\documentclass{aastex63}

\usepackage{booktabs}
\usepackage{xcolor}
\submitjournal{the Planetary Science Journal}

\shorttitle{Chemistry of Temperate Exoplanet Hazes}
\shortauthors{Moran et al.}

\begin{document}

\title{Chemistry of Temperate Super-Earth and Mini-Neptune Atmospheric Hazes from Laboratory Experiments}

\correspondingauthor{Sarah E. Moran}
\email{smoran14@jhu.edu}

\author[0000-0002-6721-3284]{Sarah E. Moran}
\affil{Department of Earth and Planetary Sciences,
 Johns Hopkins University,
Baltimore, MD 21218, USA}

\author{Sarah M. H{\"o}rst}
\affiliation{Department of Earth and Planetary Sciences,
Johns Hopkins University,
Baltimore, MD 21218, USA}
\affiliation{Hopkins Extreme Materials Institute, Johns Hopkins University, Baltimore, MD 21218, USA}
\affiliation{Space Telescope Science Institute,
Baltimore, MD 21218, USA}

\author{V{\'e}ronique Vuitton}
\affiliation{Univ. Grenoble Alpes, CNRS, CNES, IPAG, 38000 Grenoble, France}

\author{Chao He}
\affiliation{Department of Earth and Planetary Sciences,
Johns Hopkins University,
Baltimore, MD 21218, USA}

\author{Nikole K. Lewis}
\affiliation{Department of Astronomy and Carl Sagan Institute,
Cornell University, 122 Sciences Drive,
Ithaca, NY 14853, USA}

\author{Laur{\`e}ne Flandinet}
\affiliation{Univ. Grenoble Alpes, CNRS, CNES, IPAG, 38000 Grenoble, France}

\author{Julianne I. Moses}
\affiliation{Space Science Institute, Boulder, Colorado 80301, USA}

\author{Nicole North}
\affiliation{Department of Chemistry and Biochemistry,
University of Northern Iowa, Cedar Falls, IA 50614, USA}
\affiliation{Department of Chemistry and Biochemistry,
Ohio State University, Columbus, OH 43210, USA}

\author{Francois-R{\'e}gis Orthous-Daunay}
\affiliation{Univ. Grenoble Alpes, CNRS, CNES, IPAG, 38000 Grenoble, France}

\author{Joshua Sebree}
\affiliation{Department of Chemistry and Biochemistry,
University of Northern Iowa, Cedar Falls, IA 50614, USA}

\author{C{\'e}dric Wolters}
\affiliation{Univ. Grenoble Alpes, CNRS, CNES, IPAG, 38000 Grenoble, France}

\author{Eliza M.-R. Kempton}
\affiliation{Department of Astronomy, University of Maryland, College Park, Maryland 20742, USA}

\author{Mark S. Marley}
\affiliation{NASA Ames Research Center, Mountain View, California 94035, USA}

\author{Caroline V. Morley}
\affiliation{Department of Astronomy, The University of Texas at Austin, Austin, Texas 78712, USA}

\author{Jeff A. Valenti}
\affiliation{Space Telescope Science Institute,
Baltimore, MD 21218, USA}



\begin{abstract}
Very little experimental work has been done to explore the properties of photochemical hazes formed in atmospheres with very different compositions or temperatures than that of the outer solar system or of early Earth. With extrasolar planet discoveries now numbering thousands, this untapped phase space merits exploration. This study presents measured chemical properties of haze particles produced in laboratory analogues of exoplanet atmospheres. We used very high resolution mass spectrometry to measure the chemical components of solid particles produced in atmospheric chamber experiments. Many complex molecular species with general chemical formulas C$_w$H$_x$N$_y$O$_z$ were detected. We detect molecular formulas of prebiotic interest in the data, including those for the monosaccharide glyceraldehyde, a variety of amino acids and nucleotide bases, and several sugar derivatives. Additionally, the experimental exoplanetary haze analogues exhibit diverse solubility characteristics, which provide insight into the possibility of further chemical or physical alteration of photochemical hazes in super-Earth and mini-Neptune atmospheres. These exoplanet analogue particles can help us better understand chemical atmospheric processes and suggest a possible source of \textit{in situ} atmospheric prebiotic chemistry on distant worlds.

\end{abstract}

\keywords{exoplanetary atmospheres, terrestrial atmospheres, laboratory experiments}



\section{Introduction} \label{sec:intro}
Exoplanets, those planets outside our own solar system, can now be counted in the thousands thanks to past and ongoing surveys, e.g., \textit{Kepler} \citep{BoruckiKepler} and the \textit{Transiting Exoplanet Survey Satellite (TESS)} \citep{rickerTESS}. Follow-up observations of the most promising planetary targets with the \emph{Hubble Space Telescope}, \emph{Spitzer Space Telescope}, and ground-based facilities have thus far shown a wide range of atmospheric conditions. Many of these planets host atmospheres that have muted transmission spectra \citep{wakeford2019rnaas}, indicative of significant and as of yet unidentified opacity sources in their atmospheres. Either condensate clouds or photochemical hazes in these atmospheres, or some combination thereof, are compelling candidates to explain the observed spectra \citep{knutson2014featureless,kreidberg2014clouds,dragomir2015rayleigh,sing2016continuum}. As clouds and/or hazes are observed in our solar system on every world with a substantial atmosphere, the presence of such aerosols on extrasolar worlds comes as no surprise. Yet, the possibly unique compositions of these aerosols and the energetic regimes in which they are formed remain outstanding questions.

Photochemical hazes in particular can impact planetary atmospheric temperature structure \citep[e.g.,][]{zhangpluto2017}, the chemical inventory of the atmosphere and surface \citep[e.g.,][]{grundyhazeassurfacematerial}, and ultimately the habitability of worlds near and far \citep[e.g.,][]{TrainerPNAS2006,horst2012formation}. The composition of photochemical hazes will impact their spectroscopic properties, and thus their ability to absorb radiation across the electromagnetic spectrum. Haze opacity affects general energy transport and atmospheric dynamics \citep{marley2013review, hellingreview2019}, can shield the planetary surface from harmful radiation \citep{arney2017paleorange}, and affects telescope observations of exoplanets in both transmission and emission \citep[e.g.,][]{Morley2017gj436b} and reflected light \citep[e.g.,][]{gao2017sulfur}.

Titan, the largest moon of Saturn, is the best-studied hazy world of our solar system and provides critical context for the study of hazes on other worlds. However, \textit{in-situ} measurements of distant solar system worlds, such as Titan, remain challenging. Therefore, a long history of laboratory experiments has shed light on the formation, physical properties, and chemical structures of potential hazes in the atmospheres of solar system planets \citep{cabletholinreview2012}. These experiments have provided insights into the chemical pathways to haze formation in Titan's atmosphere \citep[e.g.,][]{vuitton2010HCN,bonnet2013compositional,gautier2014orbitrap,gautier2016hplcorbitrap,horst2018titanlabparticlegas}, and revealed that photochemical processes can produce amino acids and nucleobases suggestive of prebiotic chemistry \citep{horst2012formation}. This legacy of laboratory work has contributed greatly to our understanding of Titan's overall atmospheric chemistry and climate (see, e.g., \citealt{horsttitanreview}).

The haze analogues formed in these solar system experiments -- so-called ``tholins'' -- have thus far been the product mainly of methane, nitrogen, and carbon monoxide gas mixtures that represent the atmosphere of Titan or conditions on the early Earth \citep[e.g.,][]{horst2018earlyearth}. Additionally, models of exoplanet photochemistry have also primarily focused on ``hydrocarbon'' hazes similar to that of Titan \citep[e.g.,][]{howe2012theoretical,millerriccikempton2012chemistry,morley2013quantatively,morley2015thermal,kawashima2019hazemodel}, if only because these are the chemical pathways for which there are data.
Experiments exploring the wide range of possible atmospheric conditions found in exoplanet atmospheres remain mostly untapped. The few that have been performed have focused either on optical properties of essentially Titan-like atmospheres with increased oxidation to mimic early Earth-like exoplanets \citep{gavilan2017,gavilan2018organicaerosols} or gas phase chemistry of hot Jupiter-like atmospheres with temperatures in excess of 1000 K and with H$_2$/CO-dominated gas mixtures \citep{fleury2019HJexperiments}.

This work presents the first solid phase chemical composition measurements from a series of experiments designed to explore the wide range of possible atmospheric compositions for sub-Neptune planets. Current exoplanet population statistics suggest a dichotomy between planets 1.75-3.0 R$_\oplus$ and planets 1.1-1.75 R$_\oplus$ \citep{fulton2017,Fulton2018,hardegreeullman2020}, which have been termed ``super-Earths'' and ``mini-Neptunes,'' respectively. Theories of planetary formation and evolution have suggested that these could be two distinct planet classes that differ due to the presence or absence  of a substantial hydrogen-helium envelope, which is then eroded by subsequent stellar photoevaporation  \citep{lopezfortney2014,OwenWu2016,LehmerCatling2017,cloutierandmenou2019}. Another model, core-powered mass loss, suggests that these planets form with hydrogen-poor atmospheres  \citep{GuptaSchlichting2019}, and can also explain the radius gap between mini-Neptunes and super-Earths as the result of late-stage planet-disk interactions. Current population statistics do not favor one model over the other \citep{Loyd2020}, and it is unclear whether these are in fact two separate outcomes of planet formation or if they are a single planet population sculpted by atmospheric evolution through time \citep{Leconte2015}.

Moreover, observational data to determine the atmospheric compositions of these planets is also extremely sparse. Only two observational constraints at the mini-Neptune end of this planet distribution currently exist, and have confirmed hydrogen-rich atmospheres for two planets, K2-18~b \citep{benneke2019k2,tsiaras2019k2} and GJ 3470 b \citep{benneke2019h2subneptune}. On the super-Earth end of the planet distribution, while H$_2$-rich atmospheres have been ruled out for a number of planets \citep[e.g.,][]{demory2016,Kreidberg2019}, no definitive atmospheric composition constraints are possible with current instruments. Compositional constraints of heavier mean molecular weight atmospheres will require the higher-precision capabilities of future observatories like the \textit{James Webb Space Telescope}, the \textit{ARIEL Space Telescope}, or Extremely Large Telescopes on the ground. Therefore, the experiments described here have had to rely on atmospheric modeling approaches to determine the likely kind of atmospheres to consider for super-Earths and mini-Neptunes. These theoretical modeling studies have shown that these atmospheres could range from secondary ``terrestrial'' compositions due to outgassing to primordial H$_2$-dominated compositions \citep{ElkinsTanton2008super-earths,schaefer2012,moses2013equilibriumcomposition,hu2014photochemistry,fortney2013framework}.

Previous measurements resulting from these super-Earth to mini-Neptune experiments have reported production rates for a range of composition, temperature, and energy sources \citep{horst2018production,he2018hazeuvplasma}, the color and size of haze particles \citep{he2018particle}, and the gas phase chemistry occurring during the experiments \citep{he2019gasphase}. Here, we explore the effect of temperature, composition, and energy source on the chemistry of the resulting solid haze particles across the range of experimental conditions.

\section{Methods} \label{sec:style}
We produced analogue haze particles in an atmospheric chamber under theoretical super-Earth and mini-Neptune conditions. We then collected the solid sample produced in this experiment and performed very high resolution mass spectrometry with a Thermo Fisher Scientific LTQ-Orbitrap XL mass spectrometer. We also performed elemental combustion analysis to provide a starting point for the compositional study in order to identify specific molecules. Once measurements were taken, we used custom IDL software, \textit{idmol}, to analyze the data and make molecular identifications. A detailed summary of each step in our procedure follows.
\subsection{Laboratory Haze Sample Production} \label{sec:production}
\begin{figure*}[ht]
\centering
\includegraphics[angle=0,width=0.75\linewidth]{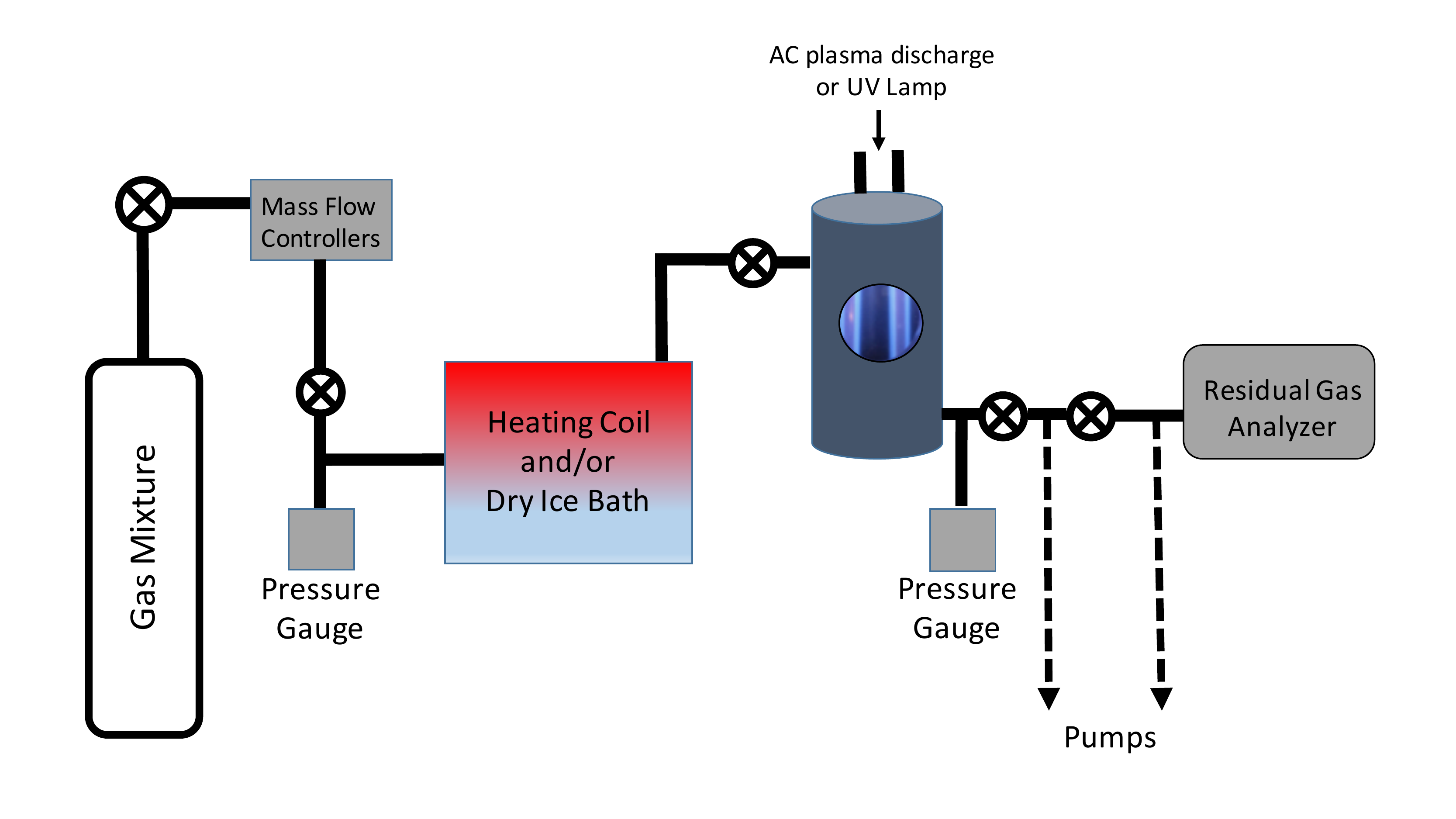}
\caption{Generalized schematic of PHAZER chamber experimental apparatus used to produce the exoplanet haze analogues. Specific gas mixtures, temperature, and energy source differs between experimental conditions.}
\label{fig:phazer}
\end{figure*}

We produced laboratory exoplanet haze analogues in the PHAZER chamber \citep{he2017co} at Johns Hopkins University, a set-up that allows us to simulate particle production over a variety of atmospheric conditions. A schematic of the PHAZER chamber and supporting equipment is provided below as Figure \ref{fig:phazer}. 

The conditions explored for this particular experiment target a broad range of possible super-Earth and mini-Neptune atmospheric conditions, including three different temperatures (300 K, 400 K, and 600 K) and two kinds of energy sources: a Lyman-$\alpha$ UV lamp, which is a proxy for the UV flux from a stellar host; and an AC cold plasma discharge. The AC cold plasma glow discharge does not directly replicate a specific atmospheric process, but it is a useful proxy for the energetic environments of planetary upper atmospheres in which dissociation of more stable molecular bonds occurs \citep{cabletholinreview2012}.

Within each temperature bin, we simulated three compositional regimes: 100$\times$, 1000$\times$, and 10000$\times$ metallicity atmospheres. Metallicity is the enhancement factor for all elements other than hydrogen and helium, relative to composition of the solar atmosphere. Broadly, our experimental conditions simulated hydrogen-rich, water-rich, and carbon-dioxide rich atmospheres at the three temperatures. These compositional breakdowns were determined through equilibrium chemistry calculations \citep{moses2013equilibriumcomposition} for each temperature at 1 mbar in atmospheric pressure. Chemical equilibrium is a good first-order approximation of the dominant available constituents in a planetary atmosphere. Various modeling approaches \citep{moses2013equilibriumcomposition,hu2014photochemistry} have suggested a range of possible gas mixtures resulting from chemical equilibrium. These cases can range from H$_2$-rich atmospheres, likely more representative of a primordial atmosphere that accreted directly from the proto-planetary disk, to outgassed atmospheres dominated by water or carbon dioxide. Without a statistically significant sample of observational constraints to work from, our gas mixtures are by necessity determined from theoretical modeling outputs. Our experimental conditions therefore sample a range of potential theoretical atmospheric outcomes thought to be common for super-Earths and mini-Neptunes. We derive the mixing ratios from equilibrium chemistry calculations based on the Chemical Equilibrium and Applications code (CEA, \citealt{gordonandmcbride}) and cap the constituent gases present at 1\% or greater to provide a reasonable amount of experimental complexity. More details about the reasoning behind our initial gas mixtures can be found in \citet[]{horst2018production,he2018particle,he2018hazeuvplasma}.

Table \ref{table:gasmixtures1} lists initial gas mixing ratios for all nine experimental conditions. Each experiment was run with gases flowing continuously for 72 hours to produce ample solid sample and to provide comparison to previous Titan experimental production rates \citep{horst2018production}. Each experiment was performed at 1 mbar in pressure, where haze formation occurs in Titan's atmosphere \citep{cabletholinreview2012,horsttitanreview} and where we perform Titan tholin experiments for comparison. The experimental chamber was then moved to a dry ($<$ 0.1 ppm H$_2$O), oxygen-free ($<$ 0.1 ppm O$_2$)  N$_2$ glove box (Inert Technology Inc., I-lab 2GB). Within the glove box, solid sample produced was collected from the chamber walls (in the case of high production) and from mica or glass discs placed at the bottom of the chamber during the experiment (in the case of low sample production). In the dry, oxygen-free glove box, samples were then transferred to plastic vials or cases, which were then sealed with parafilm and covered with aluminium foil for storage. The use of the glove box prevented alteration of the samples by ambient Earth atmospheric conditions or light sources. Additional details about the sample production can also be found in \citet[]{horst2018production,he2018particle,he2018hazeuvplasma,he2019gasphase}.

\begin{table*}[ht]
\centering
\begin{tabular}{lccccccccc}
\hline
&&&&& Metallicity &&&& \\
\hline
{Temperature} &&& 100$\times$ && 1000$\times$ && 10000$\times$ \\
\hline
600 K &&& 72.0\% H$_2$ && 42.0\% H$_2$ && 66.0\% CO$_2$ \\  
        &&& 6.3\% H$_2$O && 20.0\% CO$_2$ && 12.0\% N$_2$ \\
        &&& 3.4\% CH$_4$ && 16.0\% H$_2$O && 8.6\% H$_2$ \\
        &&& 18.3\% He && 5.1\% N$_2$ && 5.9\% H$_2$O \\
        &&&          && 1.9\% CO && 3.4\% CO \\
        &&&         && 1.7\% CH$_4$ && 4.1\% He \\
        &&&         && 13.3\% He    && \\
\hline
400 K &&& 70.0\% H$_2$ && 56.0\% H$_2$O && 67.0\% CO$_2$ \\  
            &&& 8.3\% H$_2$O && 11.0\% CH$_4$ && 15.0\% H$_2$O \\
            &&& 4.5\% CH$_4$ && 10.0\% CO$_2$ && 13.0\% N$_2$ \\
            &&& 17.2\% He && 6.4\% N$_2$ && 5.0\% He \\
            &&&         && 1.9\% H$_2$ && \\
            &&&         && 14.7\% He && \\
\hline
300 K &&& 68.6\% H$_2$ && 66.0\% H$_2$O && 67.3\% CO$_2$ \\
              &&& 8.4\% H$_2$O && 6.6\% CH$_4$ && 15.6\% H$_2$O \\
              &&& 4.5\% CH$_4$ && 6.5\% N$_2$ && 13.0\% N$_2$ \\
              &&& 1.2\% NH$_3$ && 4.9\% CO$_2$ && 4.1\% He \\
              &&& 17.3\% He && 16.0\% He && \\
\hline
\hline
PHAZER Titan ``tholin'' &&& 95.0\% N$_2$ &&&\\
                        &&& 5.0\% CH$_4$ &&& \\
                        \hline
\end{tabular}
\caption{Initial gas mixtures used in each exoplanet experiment, determined by equilibrium chemistry calculations at the specified pressure and composition relative to the Sun \citep{moses2013equilibriumcomposition}. Metallicities of 100$\times$, 1000$\times$, and 10000$\times$ solar generally correspond to H$_2$-rich, H$_2$O-rich, and CO$_2$-rich atmospheres. PHAZER Titan gas mixture also shown.}
\label{table:gasmixtures1}
\end{table*}

\subsection{Orbitrap Mass Spectrometry Measurements}
\begin{figure*}[ht]
\centering
\gridline{\fig{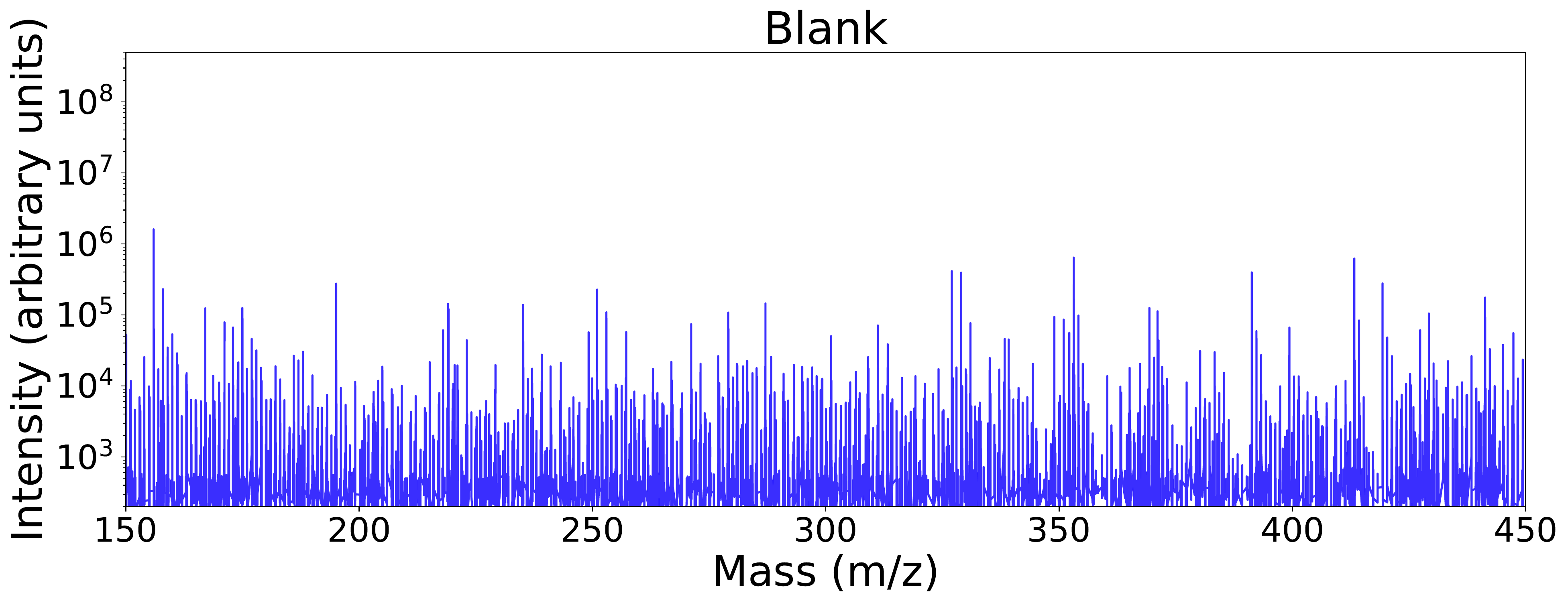}{0.49\linewidth}{}
          \fig{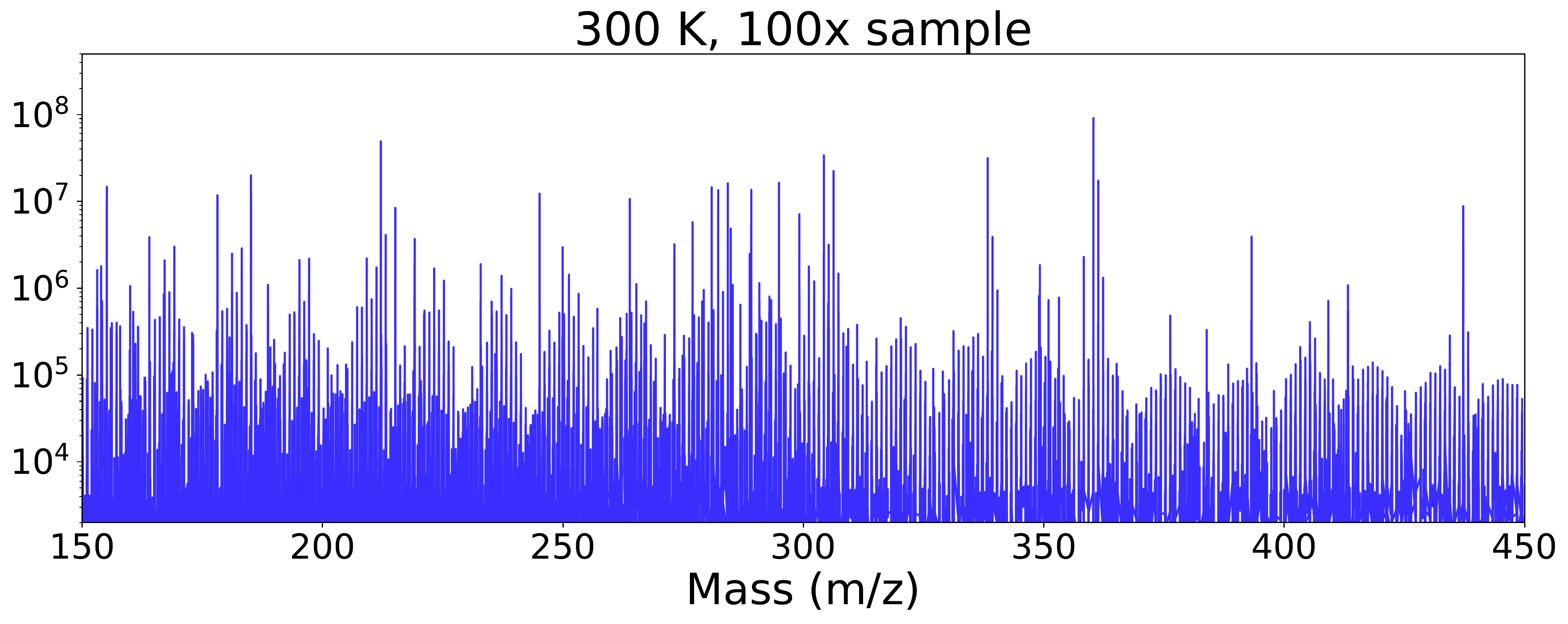}{0.49\linewidth}{}}
\caption{Mass spectrum of a blank (left). Mass spectrum of a soluble sample (right). The blank spectrum intensity is typically lower and no clear structure exists as compared to the mass spectrum of the sample. The insoluble samples have mass spectra that appear more similar to blank (left) than to the samples that were soluble (right). We also compared assigned peaks in the sample against the blank to ensure no potential contamination was unaccounted for.}
\label{fig:blank_ms}
\end{figure*}

Each sample was prepared immediately prior to performing measurements, in order to minimize contamination by ambient atmosphere. If enough solid sample was produced, we dissolved each sample in CH$_3$OH (methanol) at 1 mg/mL. If the PHAZER chamber produced only a thin film, we collected the film from the mica or glass disc by soaking the disc in 1 mL of CH$_3$OH for a minimum of 3 hours before collecting the resulting CH$_3$OH-sample mixture and transferring it into a vial. Samples then underwent sonification (1 hr) and centrifugation (5 minutes, 10000 rpm) before an additional dilution at 1 mg/mL in CH$_3$OH. The soluble fraction of the sample was then injected into a Thermo Fisher Scientific LTQ-Orbitrap XL mass spectrometer \citep{huorbitrap2005,perryorbitrap2008} with electrospray ionization (ESI) (IPAG, Grenoble, France). The Orbitrap provides high resolution mass spectrometry, with resolving power better than 10$^5$ between 200 m/z and 400 m/z and exact mass determination accuracy of $\pm$2ppm. ``Blank'' solutions from either a blank sample vial or disc and CH$_3$OH, but no sample, were also injected and measured in the Orbitrap to account for any possible background contamination in the measurements (see Figure \ref{fig:blank_ms}). Mass calibration using Thermo Fisher Scientific caffeine, MRFA peptide, and Ultramark solution was performed prior to measurements each day. Measurements were taken in three mass-to-charge (m/z) range bins, from 50 - 300 m/z, 150 - 450 m/z, and 400 - 1000 m/z. Overlap between bins ensures that signal at the edges of mass bins is properly accounted for. Instrument settings in each mass range were adjusted to ensure the best signal: the tube lens was set to 50 V, 70 V, and 90 V, respectively. We obtained 128 microscans at a flow rate of 3 $\mu$L/min with 4 scans per mass bin. We obtained measurements in both positive and negative ion polarities, as the resulting ions have displayed different molecular formulas for previous studies and thus allow a more complete view of the whole sample \citep{horstthesis,bonnet2013compositional}. Samples in solution were stored in the refrigerator when not in use. 

\begin{figure*}[ht]
\centering
\includegraphics[angle=0,width=0.95\linewidth]{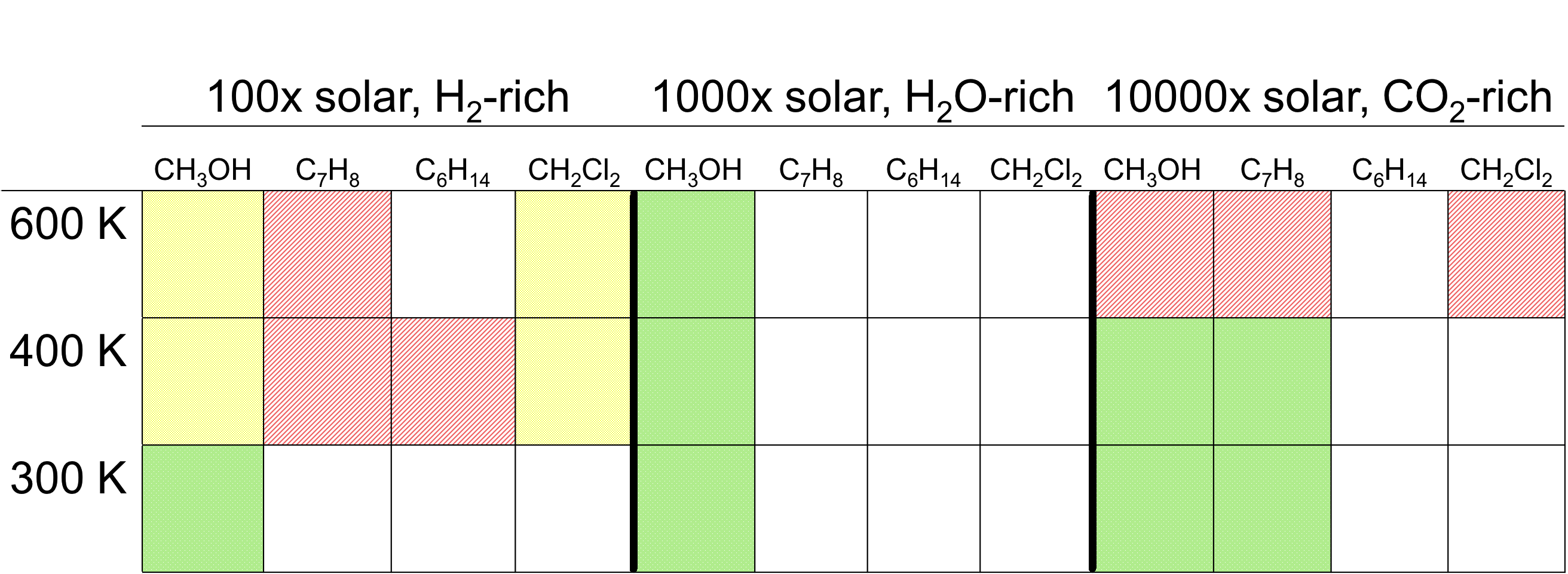}
\includegraphics[angle=0,width=0.40\linewidth]{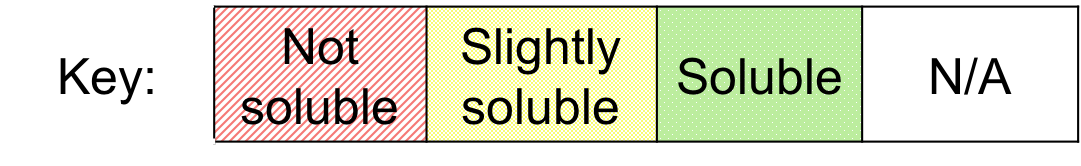}
\caption{Results of testing various solvents to dissolve the exoplanet haze analogue solid products for use in the Orbitrap. All samples here were produced by plasma discharge, as the amount of UV-produced samples tend to be small and qualitative solubility observations are not possible. Red hatched squares indicate complete lack of solubility, yellow checkered squares indicate that solids partially dissolved, and green shaded squares indicate substantial solubility. The solvents were tested in subsequent order left-to-right, stopping if a solvent dissolved the sample. The solvents tested were methanol (CH$_3$OH), followed by a toluene-methanol (C$_7$H$_8$ - CH$_3$OH) solution, followed finally by a hexane-methanol (C$_6$H$_{14}$ - CH$_3$OH) and/or a dichloromethane-methanol (CH$_2$Cl$_2$ - CH$_3$OH) solution.}
\vspace{20pt}
\label{fig:solubility}
\end{figure*}

As some samples were insoluble in CH$_3$OH, additional solvents were also used in combination with CH$_3$OH, including toluene (C$_7$H$_8$), dichloromethane (CH$_2$Cl$_2$), and hexane (C$_6$H$_{14}$). Figure \ref{fig:solubility} shows which haze analogues were dissolved in which solvents. See Section \ref{subsection:solubility} for further discussion about the solubility of the haze samples. These additional solvents were combined in approximately 1:1 solutions with methanol. Data acquisition and preliminary processing were performed with Thermo Fisher Scientific Xcalibur software provided by the manufacturer. 

\subsection{Combustion Analysis}

\begin{figure*}[ht!]
\centering
\includegraphics[angle=0,width=0.95\linewidth]{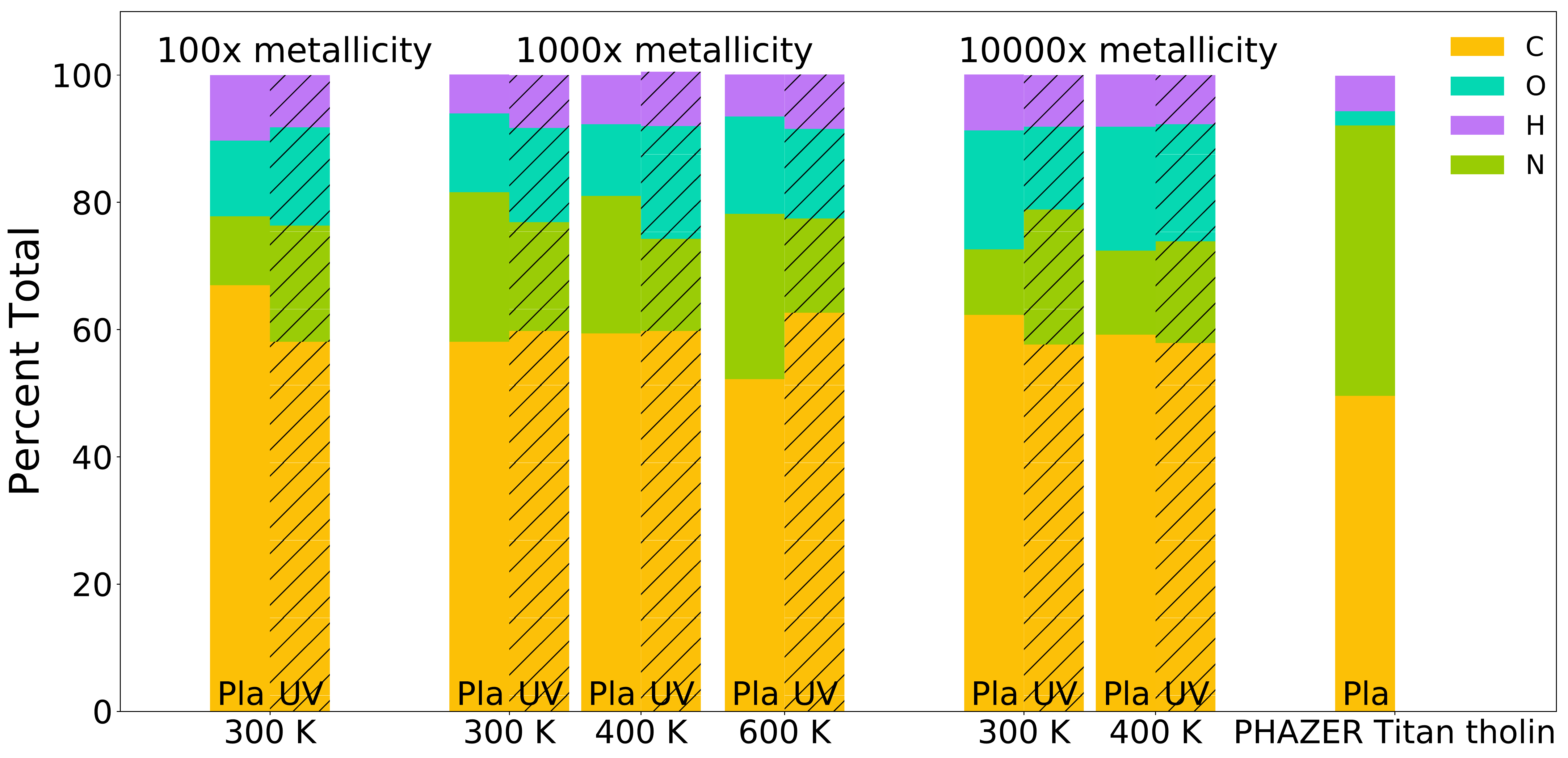}
\caption{Results of elemental analysis performed with assignments based on Orbitrap MS measurements and \textit{idmol} analysis. ``Pla'' and ``UV'' labels denote whether the sample was produced via AC plasma or the UV lamp energy source. These values are compared to PHAZER standard Titan tholin sample, with elemental ratios determined by combustion analysis. All exoplanet experimental samples have dramatically more oxygen than the Titan sample, presumably due to enhanced oxygen in the initial gas mixtures, suggesting that oxygen is readily incorporated into the solid. These measurements are subject to significant uncertainties as discussed in Section \ref{subsec:data_analysis} and reported in Tables \ref{table:elementalratiosplasma} and \ref{table:elementalratiosuv}.}
\label{fig:elemental_analysis}
\end{figure*}

Elemental combustion analysis was performed with a Thermo Scientific Flash 2000 Elemental Analyzer (Department of Chemistry and Biochemistry, University of Northern Iowa, IA, USA) on the two haze analogues that produced the most sample volume, the 400 K and 300 K at 1000$\times$ solar metallicity under the plasma source. We placed 1 to 2 mg of each sample in the analyzer for combustion analysis. The resulting elemental percentages of C, H, and N are directly measured and the percentage of O is then determined by mass subtraction. These elemental ratios are presented in Table \ref{table:elementalratiosplasma} for the plasma products and Table \ref{table:elementalratiosuv} for the UV products. PHAZER standard ``Titan tholin'' composition (produced from a 5\% CH$_4$ in N$_2$ gas mixture) is provided as a point of comparison. Figure \ref{fig:elemental_analysis} shows this information in graphic form.

\begin{table}[ht]
    \centering
    \begin{tabular}{lcccccc}
    \hline
   Sample & Carbon wt\% & Hydrogen wt\% & Nitrogen wt\% & Oxygen wt\% & C/O ratio & C/N ratio  \\
       \hline
       \hline
       \textit{via Orbitrap MS} & Plasma & Plasma & Plasma & Plasma & Plasma & Plasma \\
       \hline
       600 K, 100$\times$ & - & - & - & - & - & - \\
       400 K, 100$\times$  & - & - & - & - & - & - \\
       300 K, 100$\times$ & 67.0$\pm5.2$\% & 10.3$\pm1.0$\% & 10.8$\pm6.6$\% & 11.9$\pm6.6$\% & 5.6{\small $\pm3.2$} & 6.2$\pm3.8$ \\
       \hline
       600 K, 1000$\times$ & 52.2$\pm2.6$\% & 6.6$\pm1.7$\% & 26.0$\pm11.0$\% & 15.3$\pm8.0$\% & 3.4$\pm1.8$ & 2.0$\pm0.9$ \\
       400 K, 1000$\times$  & 59.4$\pm4.5$\% & 7.7$\pm1.1$\% & 21.6$\pm2.5$\% & 11.3$\pm3.9$\% & 5.3$\pm1.9$ & 2.7$\pm0.4$ \\
       300 K, 1000$\times$ & 58.1$\pm1.3$\% & 6.1$\pm0.4$\% & 23.5$\pm2.1$\% & 12.4$\pm2.4$\% & 4.7$\pm0.9$ & 2.5$\pm0.2$ \\
       \hline
       600 K, 10000$\times$ & - & - & - & - & - & -\\
       400 K, 10000$\times$  & 59.2$\pm4.2$\% & 8.2$\pm1.2$\% & 13.2$\pm17.0$\% & 19.5$\pm11.6$\% & 3.0$\pm1.8$ & 4.5$\pm5.8$ \\
       300 K, 10000$\times$ & 62.3$\pm5.1$\% & 8.8$\pm1.1$\% & 10.3$\pm13.2$\% & 18.7$\pm6.9$\% & 3.3$\pm1.3$ & 6.1$\pm7.8$ \\
       \hline
       \hline
       \textit{via Combustion} &&&&&&\\
       \hline
       400 K, 1000$\times$  & 56$\pm2.5$\% & 6.1$\pm0.2$\% & 21.1$\pm0.5$\% & 17$\pm3.2$\% & 3.3$\pm0.6$ & 2.7$\pm0.6$ \\
       300 K, 1000$\times$ & 51$\pm1.2$\% & 6.1$\pm0.1$\% & 27.1$\pm0.7$\% & 15$\pm2.0$\% & 3.4$\pm0.5$ & 1.9$\pm0.5$ \\
       PHAZER Titan ``tholin'' & 49.6$\pm0.5$\% & 5.6$\pm0.5$\% & 42.5$\pm0.5$\% & 2.2$\pm0.5$\% & 22.5$\pm0.5$ & 1.2$\pm0.5$ \\
      \hline
      \hline
    \end{tabular}
    \centering \caption{For samples produced by plasma discharge, elemental ratios and associated carbon-to-oxygen and carbon-to-nitrogen ratios. Some plasma samples were not soluble and thus were not subjected to further analysis; these rows are left empty in the table. Errors from the Orbitrap are the standard deviation of all mass ranges for both positive and negative ions for each sample. Errors reported for combustion analysis are the standard deviations of 3 runs for the 400 K sample and 4 runs for the 300 K sample. Similar results from the combustion analysis confirm that the \textit{idmol} molecular assignments based on LTQ Orbitrap measurements are accurate. Standard PHAZER Titan ``tholin'' elemental analysis provide a point of comparison.} 
    \label{table:elementalratiosplasma}
\end{table}

\begin{table}[ht]
    \centering
    \begin{tabular}{lcccccc}
    \hline
   Sample & Carbon wt\% & Hydrogen wt\% & Nitrogen wt\% & Oxygen wt\% & C/O ratio & C/N ratio  \\
       \hline
       \hline
       \textit{via Orbitrap MS} & UV & UV & UV & UV & UV & UV \\
       \hline
       600 K, 100$\times$ & -- & -- & -- & -- & -- & --\\
       400 K, 100$\times$  & -- & -- & -- & -- & -- & --\\
       300 K, 100$\times$ & 58.1$\pm7.3$\% & 8.2$\pm1.1$\% & 18.3$\pm5.6$\% & 15.4$\pm5.1$\% & 3.8$\pm1.3$ & 3.2$\pm1.1$ \\
       \hline
       600 K, 1000$\times$ & 62.7$\pm7.2$\% & 8.5$\pm1.2$\% & 14.8$\pm12.7$\% & 14.1$\pm4.4$\% & 4.4$\pm1.5$ & 4.3$\pm3.7$ \\
       400 K, 1000$\times$  & 59.8$\pm5.4$\% & 8.5$\pm1.5$\% & 14.5$\pm11.0$\% & 17.7$\pm3.3$\% & 3.4$\pm0.7$ & 4.1$\pm3.2$ \\
       300 K, 1000$\times$ & 59.8$\pm6.5$\% & 8.3$\pm1.1$\% & 17.1$\pm7.2$\% & 14.8$\pm4.8$\% & 4.1$\pm1.4$ & 3.5$\pm1.5$ \\
       \hline
       600 K, 10000$\times$ & -- & -- & -- & -- & -- & --\\
       400 K, 10000$\times$  & 57.9$\pm7.4$\% & 7.7$\pm0.8$\% & 16.0$\pm10.8$\% & 18.4$\pm8.5$\% & 3.1$\pm1.5$ & 3.6$\pm2.5$ \\
       300 K, 10000$\times$ & 57.7$\pm9.0$\% & 8.1$\pm1.1$\% & 21.2$\pm8.3$\% & 13.0$\pm6.9$\% & 4.4$\pm2.5$ & 2.7$\pm1.1$ \\
       \hline
       \hline
    \end{tabular}
    \centering \caption{For samples produced by UV illumination, elemental ratios and associated carbon-to-oxygen and carbon-to-nitrogen ratios. Errors reported are the standard deviation of all mass ranges and both polarities for each sample. Some plasma products were insoluble and unable to provide adequate signal for measurement and analysis. The corresponding UV samples also had very poor signal and attempts at analysis were inconclusive. Compositional differences between the samples produced by different energy sources exist, but mostly fall within error.} 
    \label{table:elementalratiosuv}
\end{table}

\subsection{Data Analysis} \label{subsec:data_analysis}

Only samples which showed substantial solubility (see Section \ref{subsection:solubility}) were subjected to detailed data analysis, i.e., the green shaded boxes in Figure \ref{fig:solubility}. We accounted for solubility and potential contamination in two ways. We compared the mass spectrum of a blank taken directly prior to the sample with the mass spectrum of the sample. The intensity of the signal in the mass spectrum was used as a first pass diagnostic; however, the Orbitrap instrument always tries to maximize the number of ions accumulated and therefore intensities alone are not sufficient to determine signal \citep{huorbitrap2005}. The next comparison was the structure of the mass spectrum itself. Repeating mass peak groupings are clearly observed in cases of true sample signal as compared to the blank, as shown in Figure \ref{fig:blank_ms}.

These data contain many hundreds to thousands of peaks, making manual identification impractical. As such, data were analyzed with custom IDL/FORTRAN software, called \textit{idmol} \citep{horstthesis}, which quickly assigns molecular peaks. First, \textit{idmol} calculates all possible molecules from the mass spectrum and then narrows down the options based on user input parameters such as the maximum number of oxygen molecules, the mass tolerance, and the nitrogen-to-carbon ratio. The program then eliminates peaks that are below the noise level or due to Fourier ringing in the most intense peaks \citep{horstthesis}. \textit{Idmol} uses the nitrogen rule (i.e., that compounds with an even nominal mass have an even number of nitrogen atoms and vice versa for compounds with odd nominal masses) to make assignments for lower mass peaks and then assigns likely higher mass peaks based upon its previous lower mass assignments. Assigned molecules are then compared against a database of known molecular formulas for prebiotic material, including amino acids, nucleobases, and simple sugars taken from the literature \citep[e.g.,][]{lufreelandaminoacidlists,cooper2018sugarreview}. Once formula assignments were made by \textit{idmol}, we checked each assigned peak in the sample against the corresponding blank to ensure any potential contamination was accounted for in the sample. No assigned peaks listed in Tables \ref{table:molecules_obshydro}, \ref{table:molecules_obswater1}, \ref{table:molecules_obswater2}, \ref{table:molecules_obswater3}, \ref{table:molecules_obswater4}, or \ref{table:molecules_obscarbon} appeared in the corresponding blank data.

Additionally, final molecular assignments from \textit{idmol} were compared to the elemental ratios from combustion analysis, as confirmation of accurate molecular identification. Elemental ratios were determined by calculating the intensity weighted average composition based on the  assignments made by \textit{idmol}. Previous work \citep{horstthesis} shows that oxygen-containing molecules tend to have lower intensities as measured by Orbitrap, and that boosting the lowest 10\% intensities by a factor 10 brings elemental analysis results from Orbitrap and combustion analysis into reasonable agreement; therefore, we have performed this same correction here. Tables \ref{table:elementalratiosplasma} and \ref{table:elementalratiosuv} and Figure \ref{fig:elemental_analysis} present results averaged over positive and negative ions. Previous analysis of Titan haze analogues demonstrates that averaging over positive and negative ion modes is necessary to obtain accurate bulk sample composition \citep{horstthesis}. Certain species are more likely to be either negatively ionized or positively ionized within the mass spectrometer, requiring measurement in both modes to describe the bulk sample. Error is reported as the standard deviation of the calculated ratios of all mass ranges for both positive and negative ions. Differing ionization efficiencies between molecules and the fact that the samples are not completely soluble will affect the Orbitrap results. Thus the intensity-weighted elemental analysis reported here has significant uncertainties associated with it, which the errors reported in Tables \ref{table:elementalratiosplasma} and \ref{table:elementalratiosuv} reflect. The elemental ratios reported here should therefore be interpreted as general trends in the bulk sample composition rather than a strict adherence to the specific values reported.

\section{Results} 
\label{sec:results}
We observe broad trends in haze chemical properties for different metallicities and temperatures, driven in part by the impact of the initial gas mixture. Further experiments isolating only temperature or only the initial gas mixtures would provide additional insight as to the particular formation conditions of each solid compound. For this work, we focus on the broad trends observed and prebiotic molecular formulas detected in each experiment.

We observe regular spacings of peak groups within each metallicity case, observing spacings of 13.5 u for the 300 K, 100$\times$ (hydrogen-rich gas mixture) hazes, between 13 and 14 u for the 1000$\times$ (water-rich gas mixture) hazes at all three temperatures, and between 10-14 u for the 10000$\times$ (carbon dioxide-rich gas mixtures) hazes at 300 and 400 K. These groupings likely correspond to chemical families in the solid products. Figures \ref{fig:ms_100x_grid}, \ref{fig:ms_1000x_grid}, and \ref{fig:ms_10000x_grid} show the mass spectrum for each sample for both positive and negative ions, as well for both plasma discharge- and UV-produced samples.

Additionally, we detect hundreds to thousands of different stoichiometries in each particular haze analogue, indicating very complex mixtures. Each individual stoichiometry represents a possible molecule. Tables \ref{table:molecules_obshydro}, \ref{table:molecules_obswater1}, \ref{table:molecules_obswater2}, \ref{table:molecules_obswater3}, \ref{table:molecules_obswater4}, and \ref{table:molecules_obscarbon} report those with the molecular formulas for amino acids (both biological and non-proteinogenic), nucleotide bases, and sugars and their derivatives for each metallicity case.

\subsection{Hydrogen-rich Atmospheres Results}

For the hydrogen-rich (100$\times$ metallicity) initial gas mixtures, only the 300 K condition produced particles that were adequately soluble for further analysis within the Orbitrap. We observe repeating mass peak group spacings of 13.5 u in both the positive and negative ions of the data, likely corresponding to additions of repeated chemical groups combining in specific ratios, as has been seen previously in studies of Titan ``tholin'' \citep{horstthesis}. In this set of gas mixtures, only the 300 K case contained NH$_3$, which suggests that ammonia, despite only being present at the $\sim$1\% level in the gas phase, plays a key role in the resulting chemical incorporation of solid particles. NH$_3$ is highly susceptible to photolysis, as demonstrated in the models of \citet{millerriccikempton2012chemistry}.

\begin{figure*}[ht]
\centering
\includegraphics[angle=0,width=0.95\linewidth]{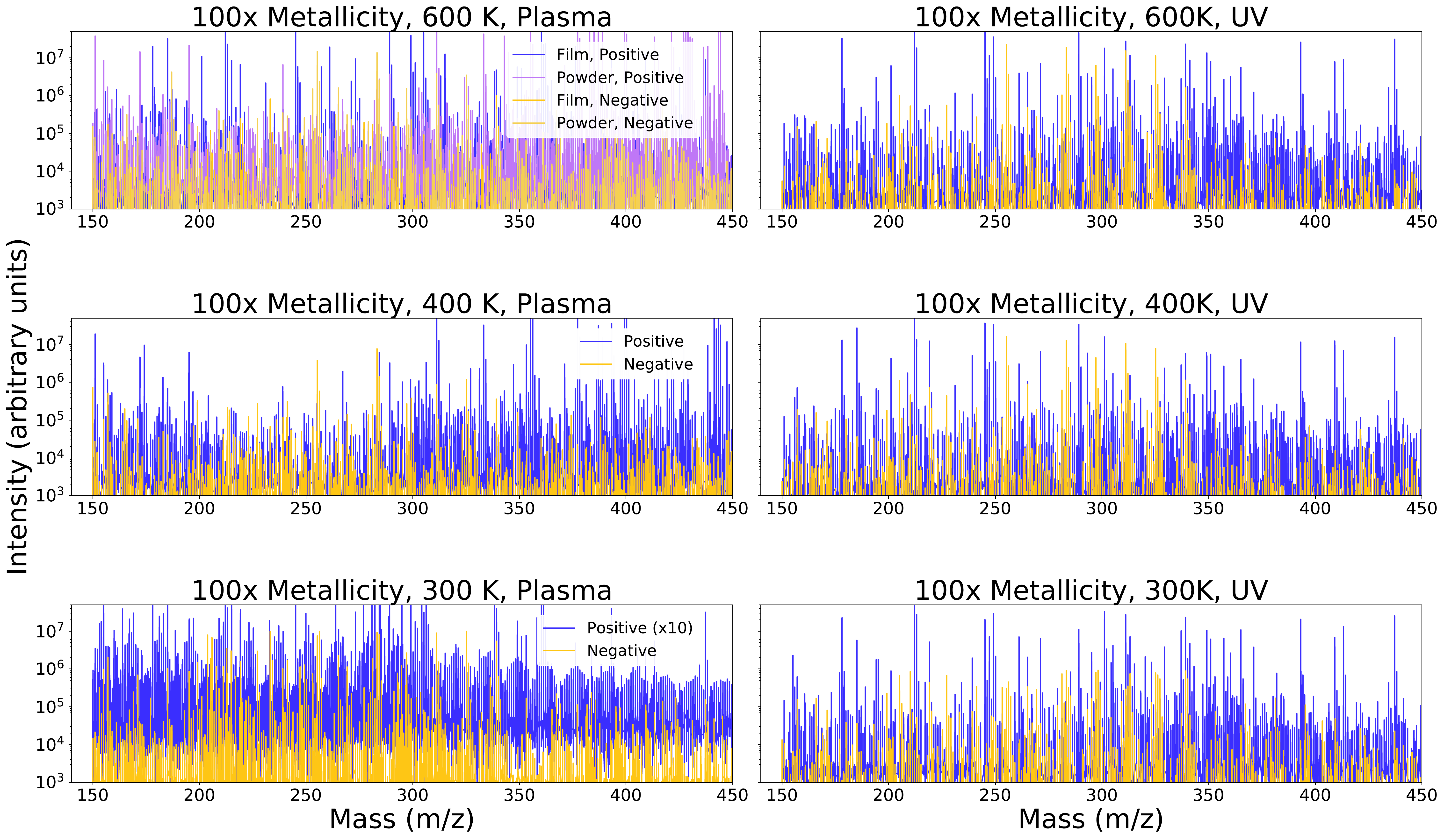}
\caption{Mass spectra from 150 to 450 m/z for all 100x metallicity plasma and UV samples, all dissolved in methanol. The 300 K plasma discharge case shows clear signs of structure, while the two higher temperature samples are noisy and were not subjected to further analysis. The 300 K plasma sample intensities were offset by a factor 10 to clearly show the stair-step structure of the mass spectra. UV sample spectra are less structured, likely due to lower sample concentrations.}
\label{fig:ms_100x_grid}
\end{figure*}

Figure \ref{fig:ms_100x_grid} shows mass spectra for all temperature cases. The 400 K and 600 K samples yielded noisy spectra with little to no structure.  Both positive and negative mode data are superimposed upon each other, showing the spectral intensity for negative ions is systematically lower than the intensity for positive ions, as is typical for the Orbitrap instrument and results from differences in ionization efficiencies between positive and negative modes and instrument systematics \citep{huorbitrap2005,perryorbitrap2008}. The 300 K plot shows intensities offset by a factor 10 so that the clearly structured stair-step pattern in the mass spectra of the positive and negative mode is more visible.

\begin{table*}[htbp]
\resizebox{\columnwidth}{!}{%
\begin{tabular}{ccccc}
\hline
\multicolumn{1}{l}{} & \multicolumn{1}{l}{} & \multicolumn{1}{l}{} & 
\multicolumn{1}{l}{100$\times$ Results} & \multicolumn{1}{l}{} \\ 
\hline\hline

\multicolumn{1}{l}{} & \multicolumn{1}{l}{} & \multicolumn{1}{l}{} & \multicolumn{1}{l}{600 K \textit{material insoluble}} & \multicolumn{1}{l}{} \\

\hline\hline

\multicolumn{1}{l}{} & \multicolumn{1}{l}{} & \multicolumn{1}{l}{} & \multicolumn{1}{l}{400 K \textit{material insoluble}} & \multicolumn{1}{l}{} \\

\hline \hline

\multicolumn{1}{l}{} & \multicolumn{1}{l}{} & \multicolumn{1}{l}{} & \multicolumn{1}{l}{300 K} & \multicolumn{1}{l}{}
\\ 
\textit{Mass (m/z) $\pm\Delta$ppm} & \textit{Detection} & \textit{Formula} & \textit{Potential Molecule} & \textit{Relevance} \\

 \begin{tabular}[c]{@{}l@{}}90.0317 $\pm2.4$\\ 129.0426 $\pm0.9$\\ 135.0545 $\pm2.1$\\ 147.0532 $\pm1.5$\\ 155.0695 $\pm2.3$\\ 159.0895 $\pm0.8$\\ 211.0845 $\pm1.3$\\219.0743 $\pm1.8$\\246.1216 $\pm1.2$\end{tabular} 

& \begin{tabular}[c]{@{}l@{}} - \\ - \\ - \\ - \\ - \\ - \\ p \\ p \\ +/p \end{tabular}

& \begin{tabular}[c]{@{}l@{}}C$_3$O$_3$H$_6$\\ C$_5$NO$_3$H$_7$\\ C$_5$N$_5$H$_5$\\ C$_5$NO$_4$H$_9$\\ C$_6$N$_3$O$_2$H$_9$\\ C$_7$NO$_3$H$_{13}$\\C$_{10}$NO$_4$H$_{13}$\\ C$_8$NO$_6$H$_{13}$\\C$_{10}$N$_2$O$_5$H$_{18}$\end{tabular} 

& \begin{tabular}[c]{@{}l@{}}glyceraldehyde\\ pyroglutamic acid\\ adenine\\ glutamic acid\\ histidine\\ L-valine, N-acetyl\\tyrosine, 3-methoxy\\\textit{O}-Succinylhomoserine\\ Boc-L-glutamine\end{tabular} & \begin{tabular}[c]{@{}l@{}}monosaccharide\\ non-proteinogenic amino acid\\ nucleotide base\\ biological amino acid\\ biological amino acid\\ non-proteinogenic amino acid\\ non-proteinogenic amino acid\\non-proteinogenic amino acid\\non-proteinogenic amino acid\end{tabular} \\ \hline \hline
\end{tabular}}
\caption{Molecular formulas detected from each 100$\times$ metallicity experiment.}
\tablenotetext{}{Detection column indicates energy source and detection polarity. Plasma (+: positive ion, -: negative ion) and UV (p: positive ion, n: negative ion). We report the smaller $\Delta$ppm between measured m/z and exact m/z when a detection was made in more than one instrument mode.}
\label{table:molecules_obshydro}
\end{table*}

For our molecular detections, we report in Table \ref{table:molecules_obshydro} only the molecular formulas for amino acids, nucleotide bases, and sugars. The 300 K 100$\times$ sample showed the presence of C$_3$H$_6$O$_3$, which is the formula for glyceraldehyde. This is the first known atmospheric experiment in the absence of liquid water, to our knowledge, to detect the molecular formula of a simple sugar from the solid products (for further context and discussion, see \ref{subsection:prebiotic}). Additionally, we detected the formulas for adenine, glutamic acid, and histidine, which are a nucleobase and proteinogenic amino acids respectively, from the 300 K hydrogen-rich gas mixture. All play vital roles in Earth-based metabolisms. We explore the implications of this further in Section \ref{subsection:prebiotic}.

\subsection{Water-rich Atmospheres Results}

\begin{figure*}[ht]
\centering
\includegraphics[angle=0,width=0.95\linewidth]{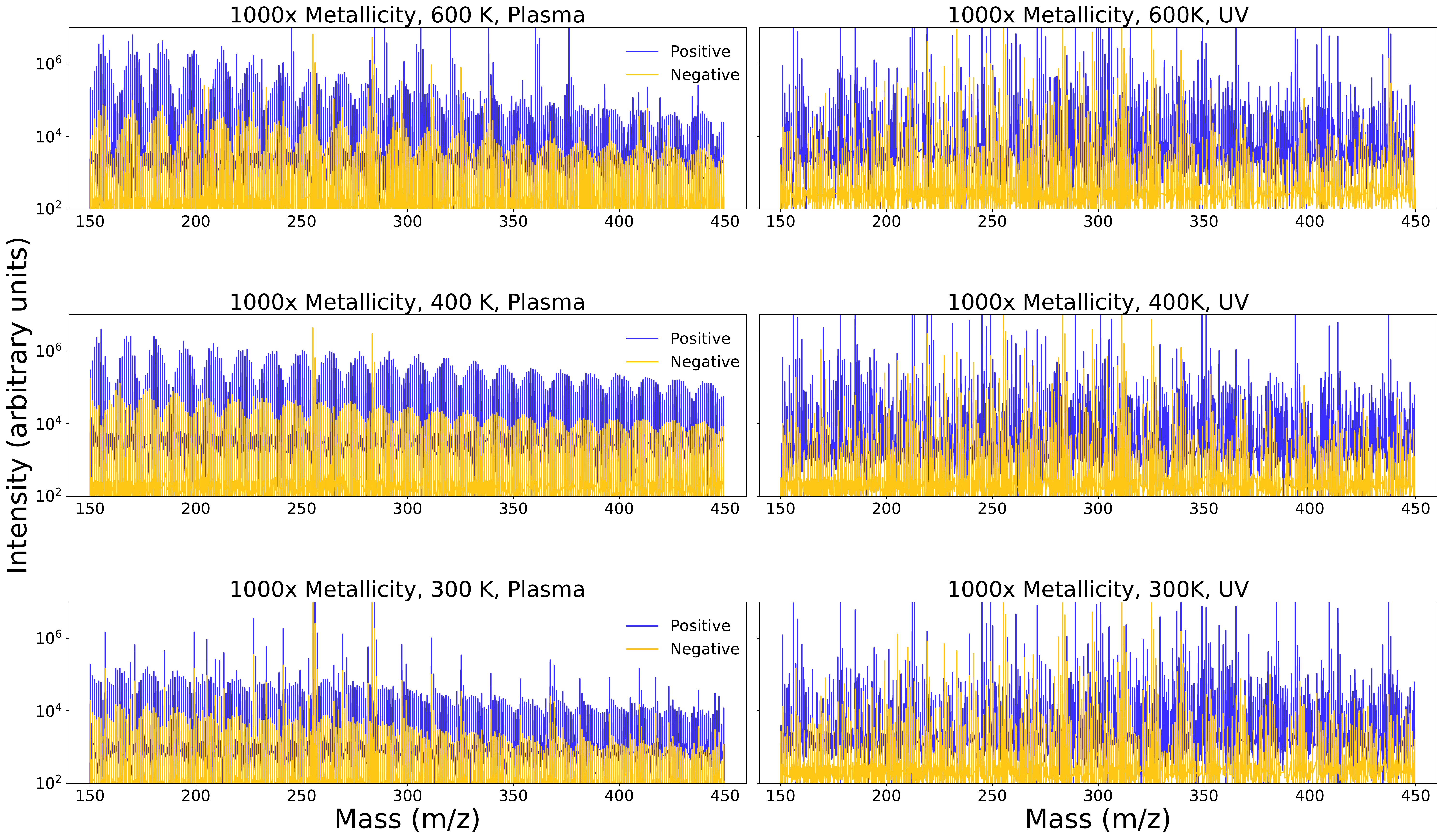}
\caption{Mass  spectra  from  150  to  450  m/z  for  all  1000$\times$ metallicity plasma  and  UV  samples, dissolved in methanol.   While all are highly structured, the 300~K case of the plasma products displays a unique shape that indicates its distinctive chemistry as compared to the hotter two samples. UV sample mass spectra are all less structured, likely due to lower sample concentrations.}
\label{fig:ms_1000x_grid}
\end{figure*}

For the water-rich (1000$\times$ metallicity) initial gas mixtures, all samples produced highly structured mass spectra, indicating that the samples are composed of highly complex molecular compounds, as well as that they were soluble in methanol. We observe repeating mass peak groupings of between 13 and 14 u in both the positive and negative ions of the data, again pointing to repeating distinct chemical groups. Mass spectra for all conditions are found in Figure \ref{fig:ms_1000x_grid}. The water-rich cases yielded the largest number of molecules with prebiotic roles, so we provide a separate table for each temperature. Tables \ref{table:molecules_obswater1} and \ref{table:molecules_obswater2} present results for 600 K, Table \ref{table:molecules_obswater3} presents results for 400 K, and Table \ref{table:molecules_obswater4} presents results for 300 K.

The 600 K and 400 K samples have similarly shaped peak groupings separated by 13 to 14 amu (averaging to 13.5 u). The 300 K sample has a different peak group shape with consistent spacing of 14 u. The water-rich cases can be differentiated by certain unique constituents: only the 600 K case contained CO and only the 300 K gas did not contain H$_2$. Additional experiments isolating changes in temperature or gas mixture could help identify the source of the 300 K sample's unique mass spectrum shape.

We detected a multitude of formulas for amino acids, both biological and non-proteinogenic, in each set of solid particles produced from the water-rich gas mixtures, for both positive and negative ions. We detected nucleotide base formulas in the water-rich samples from each set of temperatures -- all three contain the formula for guanine, while the 300 K condition additionally contains the formula for adenine. The 600 K and 300 K samples have the formula for thymine glycol, a derivative of the nucleotide base thymine. Finally, we detected the formula for the sugar acid gluconic acid in the 600 K sample and the formula for glyceraldehyde, the simplest monosaccharide, in the 300 K water-rich sample. 

\begin{table*}[ht]
\resizebox{\columnwidth}{!}{%
\begin{tabular}{lllll}
\toprule
\multicolumn{1}{l}{} & \multicolumn{1}{l}{} & \multicolumn{1}{l}{1000$\times$, 600 K Results} & \multicolumn{1}{l}{} & \multicolumn{1}{l}{}  \\ \hline
\begin{tabular}[c]{@{}l@{}}
\textit{Mass (m/z) $\pm\Delta ppm$}\\ 
102.0429 $\pm1.7$ \\
103.0633 $\pm1.7$ \\ 
114.0429 $\pm1.4$ \\ 
132.0899 $\pm1.0$ \\ 
146.0691 $\pm2.4$ \\ 
146.1055 $\pm2.3$ \\
151.0494 $\pm2.7$ \\
153.0426 $\pm2.6$ \\
153.0790 $\pm0.4$ \\
155.0695 $\pm0.02$ \\ 
156.0647 $\pm0.3$ \\ 
157.0375 $\pm3.0$ \\
157.0739 $\pm0.4$ \\
157.1103 $\pm0.6$ \\
158.0328 $\pm2.7$ \\
159.0895 $\pm3.5$ \\ 
159.1259 $\pm0.2$ \\
160.0484 $\pm3.4$ \\
160.1212 $\pm0.4$ \\
161.0688 $\pm1.9$ \\
167.0695 $\pm0.02$ \\ 
169.0851 $\pm0.13$ \\ 
171.0644 $\pm1.0$ \\ 
171.1259 $\pm0.2$ \\
172.0960 $\pm1.0$ \\ 
173.0437 $\pm2.9$ \\
174.1004 $\pm1.3$ \\ 
174.1117 $\pm0.8$ \\ 
175.0845 $\pm2.7$ \\
175.0957 $\pm0.9$ \\
181.0739 $\pm2.5$ \\
182.0804 $\pm1.5$ \\
188.1161 $\pm0.8$ \\ 
188.1273 $\pm1.1$ \\
195.0895 $\pm3.4$ \\
196.0484 $\pm3.2$ \\
196.0848 $\pm2.6$ \\
196.0583 $\pm2.2$ \\
199.0845 $\pm3.1$ \\
\end{tabular} 
\begin{tabular}[c]{@{}l@{}}
\textit{Detection}\\ 
+\\
+\\ 
+\\ 
+\\ 
+\\ 
+\\
-\\
-\\
+/-\\
+/-\\ 
+/-\\ 
-\\
+/-\\
+\\
-\\
-\\ 
+\\
-\\
+\\
- \\
+/-\\ 
+/-\\ 
+/-\\ 
+\\
+/-\\ 
-\\
+/-\\ 
+\\ 
-\\
+/-\\
-\\
+/-\\
+/-\\ 
+\\
-/n\\
-\\
-\\
+\\
-\\
\end{tabular} &
 \begin{tabular}[c]{@{}l@{}}
\textit{Formula}\\ 
C$_3$N$_2$O$_2$H$_6$\\ 
C$_4$NO$_2$H$_{9}$ \\
C$_4$N$_2$O$_2$H$_6$\\ 
C$_5$N$_2$O$_2$H$_{12}$\\ 
C$_5$N$_2$O$_3$H$_{10}$\\
C$_6$N$_2$O$_2$H$_{14}$\\
C$_5$N$_5$OH$_5$ \\
C$_7$NO$_3$H$_7$\\
C$_8$NO$_2$H$_{11}$\\
C$_6$N$_3$O$_2$H$_{9}$\\
C$_5$N$_4$O$_2$H$_8$\\
C$_6$NO$_4$H$_{7}$ \\
C$_7$NO$_3$H$_{11}$ \\
C$_8$NO$_2$H$_{15}$\\
C$_5$N$_2$O$_4$H$_{6}$\\
C$_7$NO$_3$H$_{13}$\\
C$_8$NO$_2$H$_{17}$\\
C$_5$N$_2$O$_4$H$_{8}$\\ 
C$_7$N$_2$O$_2$H$_{16}$\\ 
C$_6$NO$_4$H$_{11}$\\ 
C$_7$N$_3$O$_2$H$_{9}$\\ 
C$_7$N$_3$O$_2$H$_{11}$\\
C$_6$N$_3$O$_3$H$_{9}$\\
C$_9$NO$_2$H$_{17}$\\
C$_6$N$_4$O$_2$H$_{12}$ \\ 
C$_5$N$_3$O$_4$H$_{7}$ \\
C$_7$N$_2$O$_3$H$_{14}$ \\
C$_6$N$_4$O$_2$H$_{14}$\\
C$_7$NO$_4$H$_{13}$\\
C$_6$N$_3$O$_3$H$_{13}$\\
C$_9$NO$_3$H$_{11}$ \\
C$_7$N$_4$O$_2$H$_{10}$\\
C$_8$N$_2$O$_3$H$_{16}$\\
C$_7$N$_4$O$_2$H$_{16}$\\
C$_{10}$NO$_3$H$_{13}$ \\
C$_{8}$N$_2$O$_4$H$_{8}$ \\
C$_{9}$N$_2$O$_3$H$_{12}$ \\
C$_6$O$_7$H$_{12}$ \\
C$_{9}$NO$_4$H$_{13}$ \\
\end{tabular} & \begin{tabular}[c]{@{}l@{}}
\textit{Potential Molecule}\\ 
cycloserine \\
N,N-Dimethylglycine \\
$\beta$-cyanoalanine \\
ornithine \\
glutamine \\
lysine \\
guanine \\
\textit{p}-Aminosalicyclic acid \\
dopamine \\
histidine \\
1,2,4-Triazole-3-alanine \\
aminohexa-dienedioic acid\\
furanomycin \\
cyclohexylglycine \\
dihydroorotic acid \\
L-valine, N-acetyl \\
octanoic acid, 8-amino- \\
thymine glycol\\
L-Lysine, N$^6$-methyl-\\
2-Aminohexanedioic acid \\
$\beta$-Pyrazinyl-L-alanine \\
3-Methylhistidine \\
$\beta$-hydroxyhistidine \\
cyclohexylalanine \\
enduacididine \\
azaserine \\
formyllysine \\
arginine \\
spermidic acid\\
citrulline \\
tyrosine \\
lathyrine \\
leucine, glycyl- \\
homoarginine \\
tyrosine, O-methyl \\
phenylglycine, m-nitro \\
pyridinylmethylserine \\
gluconic acid\\
anticapsin\\
\end{tabular} & \begin{tabular}[c]{@{}l@{}}
\textit{Relevance}\\ 
non-proteinogenic amino acid\\
amino acid derivative\\
non-proteinogenic amino acid\\
non-proteinogenic amino acid \\
biological amino acid \\
biological amino acid \\
nucleotide base \\
aminobenzoic acid \\
non-proteinogenic amino acid\\
biological amino acid \\
non-proteinogenic amino acid\\
non-proteinogenic amino acid \\
non-proteinogenic amino acid \\
non-proteinogenic amino acid \\
pyrimidinemonocarboxylic acid\\
non-proteinogenic amino acid \\
non-proteinogenic amino acid \\
nucleotide base derivative \\
non-proteinogenic amino acid \\
non-proteinogenic amino acid \\
non-proteinogenic amino acid \\ 
non-proteinogenic amino acid \\ 
non-proteinogenic amino acid \\
non-proteinogenic amino acid \\
non-proteinogenic amino acid \\
non-proteinogenic amino acid \\
non-proteinogenic amino acid 
\\
biological amino acid \\
non-proteinogenic amino acid\\
non-proteinogenic amino acid\\
biological amino acid \\
non-proteinogenic amino acid\\ 
non-proteinogenic amino acid\\ 
non-proteinogenic amino acid \\
non-proteinogenic amino acid \\
non-proteinogenic amino acid \\ 
non-proteinogenic amino acid \\ 
sugar acid \\
non-proteinogenic amino acid \\  
 \end{tabular} \\
\hline\hline
\end{tabular}}
\caption{Molecular formulas detected from the 600 K, 1000$\times$ metallicity experiment.}
\tablenotetext{}{Detection column indicates energy source and detection polarity. Plasma (+: positive ion, -: negative ion) and UV (p: positive ion, n: negative ion). We report the smaller $\Delta$ppm between measured m/z and exact m/z when a detection was made in more than one instrument mode.}
\label{table:molecules_obswater1}
\end{table*}

\begin{table*}[ht]
\resizebox{\columnwidth}{!}{%
\begin{tabular}{lcccc}
\toprule
\multicolumn{1}{l}{} & \multicolumn{1}{l}{} & 
\multicolumn{1}{l}{} 
& \multicolumn{1}{l}{1000$\times$, 600 K Results} & \multicolumn{1}{l}{}
 \\ \hline
\begin{tabular}[c]{@{}l@{}}
\textit{Mass (m/z) $\pm \Delta ppm$}\\
205.0851 $\pm3.0$ \\
206.0804 $\pm3.5$ \\
208.0848 $\pm2.8$ \\
210.0641 $\pm2.8$ \\
211.0845 $\pm3.6$ \\
224.0797 $\pm3.2$ \\
226.1066 $\pm3.1$ \\ 
246.1328 $\pm1.8$ \\
255.1583 $\pm1.1$ \\
267.1219 $\pm3.0$ \\
270.0964 $\pm3.1$ \\
342.1162 $\pm1.3$ \\
465.3090 $\pm1.2$ \\
\end{tabular} &
\begin{tabular}[c]{@{}l@{}}
\textit{Detection}\\
-\\
-\\
-\\
- \\
-\\
-\\
-\\ 
+/-\\
+/-\\
-\\
-\\
n\\
+ \\

\end{tabular} &
\begin{tabular}[c]{@{}l@{}}
\textit{Formula}\\
C$_{10}$N$_3$O$_2$H$_{11}$ \\
C$_{9}$N$_4$O$_2$H$_{10}$ \\
C$_{10}$N$_2$O$_3$H$_{12}$ \\
C$_{9}$N$_2$O$_4$H$_{10}$ \\
C$_{10}$NO$_4$H$_{13}$ \\
C$_{10}$N$_2$O$_4$H$_{12}$\\
C$_9$N$_4$O$_3$H$_{14}$\\
C$_9$N$_4$O$_4$H$_{18}$\\
C$_{12}$N$_3$O$_3$H$_{21}$\\
C$_{12}$N$_3$O$_4$H$_{17}$\\
C$_{10}$N$_4$O$_5$H$_{14}$\\
C$_{12}$O$_{11}$H$_{22}$\\
C$_{26}$NO$_6$H$_{43}$ \\
\end{tabular}& \begin{tabular}[c]{@{}l@{}}
\textit{Potential Molecule}\\
tryptazan\\
benzotriazolylalanine \\
phenylasparagine \\
p-Nitrophenylalanine \\
tyrosine, 3-methoxy \\
3-hydroxykynurenine\\
alanylhistidine\\
octopine \\
pyrrolysine\\
agaritine \\
histidine, $\beta$-aspartyl \\
sucrose\\
glycocholic acid
\end{tabular}
& \begin{tabular}[c]{@{}l@{}}
\textit{Relevance}\\
non-proteinogenic amino acid\\
non-proteinogenic amino acid\\
non-proteinogenic amino acid\\
non-proteinogenic amino acid\\
non-proteinogenic amino acid\\
amino acid metabolite \\
amino acid metabolite \\
amino acid derivative \\
biological amino acid \\
non-proteinogenic amino acid \\
non-proteinogenic amino acid \\
disaccharide\\
bile acid \\
\end{tabular} \\
\hline\hline
\end{tabular}}
\caption{Molecular formulas detected (continued) from the 600 K, 1000$\times$ metallicity experiment.}
\tablenotetext{}{Detection column indicates energy source and detection polarity. Plasma (+: positive ion, -: negative ion) and UV (p: positive ion, n: negative ion). We report the smaller $\Delta$ppm between measured m/z and exact m/z when a detection was made in more than one instrument mode.}
\label{table:molecules_obswater2}
\end{table*}

\begin{table*}[ht]
\resizebox{\columnwidth}{!}{%
\begin{tabular}{lcccc}
\toprule
\multicolumn{1}{l}{} & \multicolumn{1}{l}{} & \multicolumn{1}{l}{} & \multicolumn{1}{l}{1000$\times$, 400 K Results} & \multicolumn{1}{l}{}  \\ \hline
\begin{tabular}[c]{@{}l@{}}
\textit{Mass (m/z) $\pm\Delta ppm$}\\ 
151.0494 $\pm3.9$ \\
151.0633 $\pm3.6$  \\
153.0426 $\pm1.7$ \\
153.0790 $\pm4.3$ \\
155.0695 $\pm4.4$ \\
156.0647 $\pm4.0$ \\
157.0739 $\pm3.4$ \\
159.0895 $\pm3.5$ \\
165.0790 $\pm3.5$ \\
167.0695 $\pm4.1$ \\
169.0851 $\pm0.1$ \\
171.0644 $\pm3.7$ \\
176.0586 $\pm3.7$ \\
179.0946 $\pm4.5$  \\
181.0739 $\pm3.7$ \\
182.0804 $\pm3.5$ \\
193.0739 $\pm4.2$ \\
195.0895 $\pm3.4$ \\
196.0848 $\pm3.9$ \\
204.0899 $\pm4.3$ \\
205.0851 $\pm4.4$  \\
206.0804 $\pm3.5$  \\
208.0848 $\pm4.2$ \\
210.0651 $\pm4.2$ \\
220.0848 $\pm3.8$ \\
224.0797 $\pm4.7$ \\
226.1066 $\pm4.5$ \\
236.0797 $\pm3.8$ \\
246.1004 $\pm4.0$ \\
246.1216 $\pm3.8$  \\
246.1328 $\pm3.8$  \\
255.1583 $\pm0.1$ \\
267.1219 $\pm3.0$ \\
276.1321 $\pm2.4$  \\
342.1162 $\pm1.3$ \\
449.3141 $\pm0.1$  \\
465.3090 $\pm1.3$ \\
\end{tabular} &
\begin{tabular}[c]{@{}l@{}}
\textit{Detection}\\ 
-\\
-\\
-/n\\
-\\
-\\
-\\
-\\
- \\
-\\
-\\
+/-\\
-\\
-\\
-\\
-\\
- \\
-\\
-/n\\
-\\
-\\
-\\
-\\
-\\
-\\
-\\
-\\
-\\
-\\
-\\
-\\
-\\
+/-\\
-\\
- \\
n\\
+/-\\
+\\
\end{tabular} &

\begin{tabular}[c]{@{}l@{}}
\textit{Formula}\\
C$_5$N$_5$OH$_5$\\
C$_8$NO$_2$H$_9$\\
C$_7$NO$_3$H$_7$\\
C$_8$NO$_2$H$_{11}$\\
C$_6$N$_3$O$_2$H$_9$\\
C$_5$N$_4$O$_2$H$_8$\\
C$_7$NO$_3$H$_{11}$\\
C$_7$NO$_3$H$_{13}$\\
C$_9$NO$_2$H$_{11}$\\
C$_7$N$_3$O$_2$H$_{9}$\\
C$_7$N$_3$O$_2$H$_{11}$\\
C$_6$N$_3$O$_3$H$_9$\\
C$_9$N$_2$O$_2$H$_8$\\
C$_{10}$NO$_2$H$_{13}$\\
C$_9$NO$_3$H$_{11}$\\
C$_7$N$_4$O$_2$H$_{10}$\\
C$_{10}$NO$_3$H$_{11}$\\
C$_{10}$NO$_3$H$_{13}$\\
C$_{9}$N$_2$O$_3$H$_{12}$\\
C$_{11}$N$_2$O$_2$H$_{12}$\\
C$_{10}$N$_3$O$_2$H$_{11}$\\
C$_{9}$N$_4$O$_2$H$_{10}$\\
C$_{10}$N$_2$O$_3$H$_{12}$\\
C$_{9}$N$_2$O$_4$H$_{10}$\\
C$_{11}$N$_2$O$_3$H$_{12}$\\
C$_{10}$N$_2$O$_4$H$_{12}$\\
C$_{9}$N$_4$O$_3$H$_{14}$\\
C$_{11}$N$_2$O$_4$H$_{12}$\\
C$_{13}$N$_2$O$_3$H$_{14}$\\
C$_{10}$N$_2$O$_5$H$_{18}$\\
C$_{9}$N$_4$O$_4$H$_{18}$\\
C$_{12}$N$_3$O$_3$H$_{21}$\\
C$_{12}$N$_3$O$_4$H$_{17}$\\
C$_{11}$N$_2$O$_6$H$_{20}$\\
C$_{12}$O$_{11}$H$_{22}$\\
C$_{26}$NO$_5$H$_{43}$\\
C$_{26}$NO$_6$H$_{43}$\\
\end{tabular} 
& \begin{tabular}[c]{@{}l@{}}
\textit{Potential Molecule}\\
guanine\\
2-Phenylglycine\\
\textit{p}-Aminosalicyclic acid\\
dopamine\\
histidine\\
1,2,4-Triazole-3-alanine\\
furanomycin\\
L-valine, N-acetyl\\
phenylalanine\\
$\beta$-Pyrazinyl-L-alanine\\
3-Methylhistidine\\
$\beta$-hydroxyhistidine\\
phenylglycine, m-cyano \\
homophenylalanine \\
tyrosine\\
lathyrine\\
phenylglycine, m-acetyl \\
tyrosine, O-methyl\\
pyridinylmethylserine \\
tryptophan \\
tryptazan \\
benzotriazolylalanine \\
phenylasparagine\\
\textit{p}-Nitrophenylalanine \\
5-Hydroxytryptophan \\
3-hydroxykynurenine\\
alanylhistidine \\
N-formylkynurenine \\
acetyltryptophan\\
Boc-L-glutamine \\
octopine\\
pyrrolysine\\
agaritine \\
saccharopine\\
sucrose\\
glycodeoxycholic acid \\
glycocholic acid \\
\end{tabular} 
& \begin{tabular}[c]{@{}l@{}}
\textit{Relevance}\\
nucleotide base\\
non-proteinogenic amino acid\\
aminobenzoic acid\\
non-proteinogenic amino acid\\
biological amino acid\\
non-proteinogenic amino acid\\
non-proteinogenic amino acid\\
non-proteinogenic amino acid\\
non-proteinogenic amino acid\\
non-proteinogenic amino acid\\
non-proteinogenic amino acid\\
non-proteinogenic amino acid\\
non-proteinogenic amino acid\\
non-proteinogenic amino acid\\
biological amino acid\\
non-proteinogenic amino acid\\
non-proteinogenic amino acid\\
non-proteinogenic amino acid\\
non-proteinogenic amino acid\\ 
biological amino acid \\
non-proteinogenic amino acid\\
non-proteinogenic amino acid\\
non-proteinogenic amino acid\\
non-proteinogenic amino acid\\
biological amino acid derivative \\
biological amino acid metabolite\\
biological amino acid metabolite\\
non-proteinogenic amino acid\\
biological amino acid derivative \\
non-proteinogenic amino acid \\
biological amino acid derivative \\
biological amino acid \\
non-proteinogenic amino acid\\
biological amino acid derivative\\
disaccharide\\
bile acid \\
bile acid \\
\end{tabular} \\
\hline\hline
\end{tabular}}
\caption{Molecular formulas detected from the 400 K, 1000$\times$ metallicity experiment.}
\tablenotetext{}{Detection column indicates energy source and detection polarity. Plasma (+: positive ion, -: negative ion) and UV (p: positive ion, n: negative ion). We report the smaller $\Delta$ppm between measured m/z and exact m/z when a detection was made in more than one instrument mode.}
\label{table:molecules_obswater3}
\end{table*}

\begin{table*}[ht]
\resizebox{\columnwidth}{!}{%
\begin{tabular}{lcccc}
\toprule
\multicolumn{1}{l}{} & \multicolumn{1}{l}{} & \multicolumn{1}{l}{} & \multicolumn{1}{l}{1000$\times$, 300 K Results} & \multicolumn{1}{l}{}  \\ \hline
\begin{tabular}[c]{@{}l@{}}
\textit{Mass (m/z) $\pm \Delta ppm$}\\ 
90.0317 $\pm4.5$ \\
135.0545 $\pm5.8$ \\
137.0477 $\pm5.9$\\
141.0426 $\pm4.9$ \\
151.0494 $\pm4.9$ \\
151.0633 $\pm4.7$ \\
153.0426 $\pm5.9$ \\
153.0790 $\pm5.4$ \\
155.0695 $\pm4.4$ \\
156.0647 $\pm5.1$ \\
157.0739 $\pm4.6$ \\
160.0484 $\pm4.6$ \\
165.0790 $\pm4.7$ \\
167.0695 $\pm5.4$ \\
169.0851 $\pm4.6$ \\
171.0644 $\pm3.7$ \\
176.0586 $\pm3.7$ \\
179.0946 $\pm4.5$ \\
181.0739 $\pm3.7$ \\
182.0804 $\pm3.5$ \\
193.0739 $\pm9.1$ \\
195.0895 $\pm3.4$ \\
196.0484 $\pm8.4$ \\
196.0848 $\pm2.6$ \\
204.0899 $\pm3.0$ \\
205.0851 $\pm4.4$ \\
206.0804 $\pm3.5$ \\
208.0848 $\pm1.5$ \\
210.0641 $\pm7.7$ \\
211.0845 $\pm0.1$ \\
220.0848 $\pm8.9$ \\
224.0797 $\pm8.3$ \\
226.1066 $\pm1.7$ \\
246.1004 $\pm8.1$ \\
246.1216 $\pm1.2$ \\
267.1219 $\pm6.0$ \\
276.1321 $\pm4.0$ \\
\end{tabular} &

\begin{tabular}[c]{@{}l@{}}
\textit{Detection}\\ 
-\\
-\\
-\\
-\\
+/-\\
-\\
-\\
-\\
-\\
-\\
-\\
-\\
-\\
+/-\\
+/-\\
+/-\\
+/-\\
-\\
+/-\\
+/-\\
+\\
+/-\\
+\\
+/-\\
+/-\\
+/-\\
+/-\\
+/-\\
+\\ 
p  \\
+/-\\
+\\
+/-\\
+\\ 
p  \\
+/-\\
-\\
\end{tabular} &

\begin{tabular}[c]{@{}l@{}}
\textit{Formula}\\
C$_3$O$_3$H$_6$\\
C$_5$N$_5$H$_5$\\
C$_7$NO$_2$H$_7$\\
C$_6$NO$_3$H$_7$\\
C$_5$N$_5$OH$_5$\\
C$_8$NO$_2$H$_9$\\
C$_7$NO$_3$H$_7$\\
C$_8$NO$_2$H$_{11}$\\
C$_6$N$_3$O$_2$H$_{9}$\\
C$_5$N$_4$O$_2$H$_{8}$\\
C$_7$NO$_3$H$_{11}$\\
C$_5$N$_2$O$_4$H$_{8}$\\
C$_9$NO$_2$H$_{11}$\\
C$_7$N$_3$O$_2$H$_{9}$\\
C$_7$N$_3$O$_2$H$_{11}$\\
C$_6$N$_3$O$_3$H$_9$\\
C$_9$N$_2$O$_2$H$_{8}$\\
C$_{10}$NO$_2$H$_{13}$\\
C$_{9}$NO$_3$H$_{11}$\\
C$_{7}$N$_4$O$_2$H$_{10}$\\
C$_{10}$NO$_3$H$_{11}$\\
C$_{10}$NO$_3$H$_{13}$\\
C$_{8}$N$_2$O$_4$H$_{8}$\\
C$_{9}$N$_2$O$_3$H$_{12}$\\
C$_{11}$N$_2$O$_2$H$_{12}$\\
C$_{10}$N$_3$O$_2$H$_{11}$\\
C$_{9}$N$_4$O$_2$H$_{10}$\\
C$_{10}$N$_2$O$_3$H$_{12}$\\
C$_{9}$N$_2$O$_4$H$_{10}$\\
C$_{10}$NO$_4$H$_{13}$\\
C$_{11}$N$_2$O$_3$H$_{12}$\\
C$_{10}$N$_2$O$_4$H$_{12}$\\
C$_{9}$N$_4$O$_3$H$_{14}$\\
C$_{13}$N$_2$O$_3$H$_{14}$\\
C$_{10}$N$_2$O$_5$H$_{18}$\\
C$_{12}$N$_3$O$_4$H$_{17}$\\
C$_{11}$N$_2$O$_6$H$_{20}$\\
\end{tabular} 
& \begin{tabular}[c]{@{}l@{}}
\textit{Potential Molecule}\\
glyceraldehyde\\
adenine \\
homarine \\
aminomuconic semialdehyde \\
guanine\\
2-Phenylglycine\\
\textit{p}-Aminosalicyclic acid\\
dopamine \\  
histidine\\
1,2,4-Triazole-3-alanine \\
furanomycin\\
thymine glycol \\
phenylalanine\\
$\beta$-Pyrazinyl-L-alanine\\
3-Methylhistidine\\ 
$\beta$-hydroxyhistidine\\
phenylglycine, m-cyano\\ 
homophenylalanine\\
tyrosine\\
lathyrine\\
phenylglycine, m-acetyl\\
tyrosine, O-methyl\\
phenylglycine, m-nitro\\
pyridinylmethylserine\\
tryptophan \\
tryptazan \\
benzotriazolylalanine \\
phenylasparagine\\
\textit{p}-Nitrophenylalanine \\
tyrosine, 3-methoxy\\
5-Hydroxytryptophan \\
3-hydroxykynurenine\\
alanylhistidine\\
acetyltryptophan\\
Boc-L-glutamine\\
agaritine\\
saccharopine\\
\end{tabular} 
& \begin{tabular}[c]{@{}l@{}}
\textit{Relevance}\\
monosaccharide \\
nucleotide base \\
non-proteinogenic amino acid\\
biological amino acid metabolite\\
nucleotide base \\
non-proteinogenic amino acid\\
aminobenzoic acid \\
non-proteinogenic amio acid\\
biological amino acid \\
non-proteinogenic amio acid\\
non-proteinogenic amio acid\\
nucleotide base derivative\\
non-proteinogenic amio acid\\
non-proteinogenic amio acid\\
non-proteinogenic amio acid\\
non-proteinogenic amio acid\\
non-proteinogenic amio acid\\
non-proteinogenic amio acid\\
biological amino acid\\
non-proteinogenic amio acid\\
non-proteinogenic amio acid\\
non-proteinogenic amio acid\\
non-proteinogenic amio acid\\
non-proteinogenic amio acid\\
biological amino acid \\
non-proteinogenic amino acid\\
non-proteinogenic amino acid\\
non-proteinogenic amino acid\\
non-proteinogenic amino acid\\
non-proteinogenic amino acid\\
biological amino acid derivative \\
biological amino acid metabolite \\
biological amino acid metabolite \\
biological amino acid derivative \\
non-proteinogenic amino acid\\
non-proteinogenic amino acid\\
biological amino acid derivative \\
\end{tabular} \\
\hline\hline
\end{tabular}}
\caption{Molecular formulas detected from the 300 K, 1000$\times$ metallicity experiment.}
\tablenotetext{}{Detection column indicates energy source and detection polarity. Plasma (+: positive ion, -: negative ion) and UV (p: positive ion, n: negative ion). We report the smaller $\Delta$ppm between measured m/z and exact m/z when a detection was made in more than one instrument mode.}
\label{table:molecules_obswater4}
\end{table*}

\subsection{Carbon Dioxide-rich Atmospheres Results}

\begin{figure*}[h!]
\centering
\includegraphics[angle=0,width=0.95\linewidth]{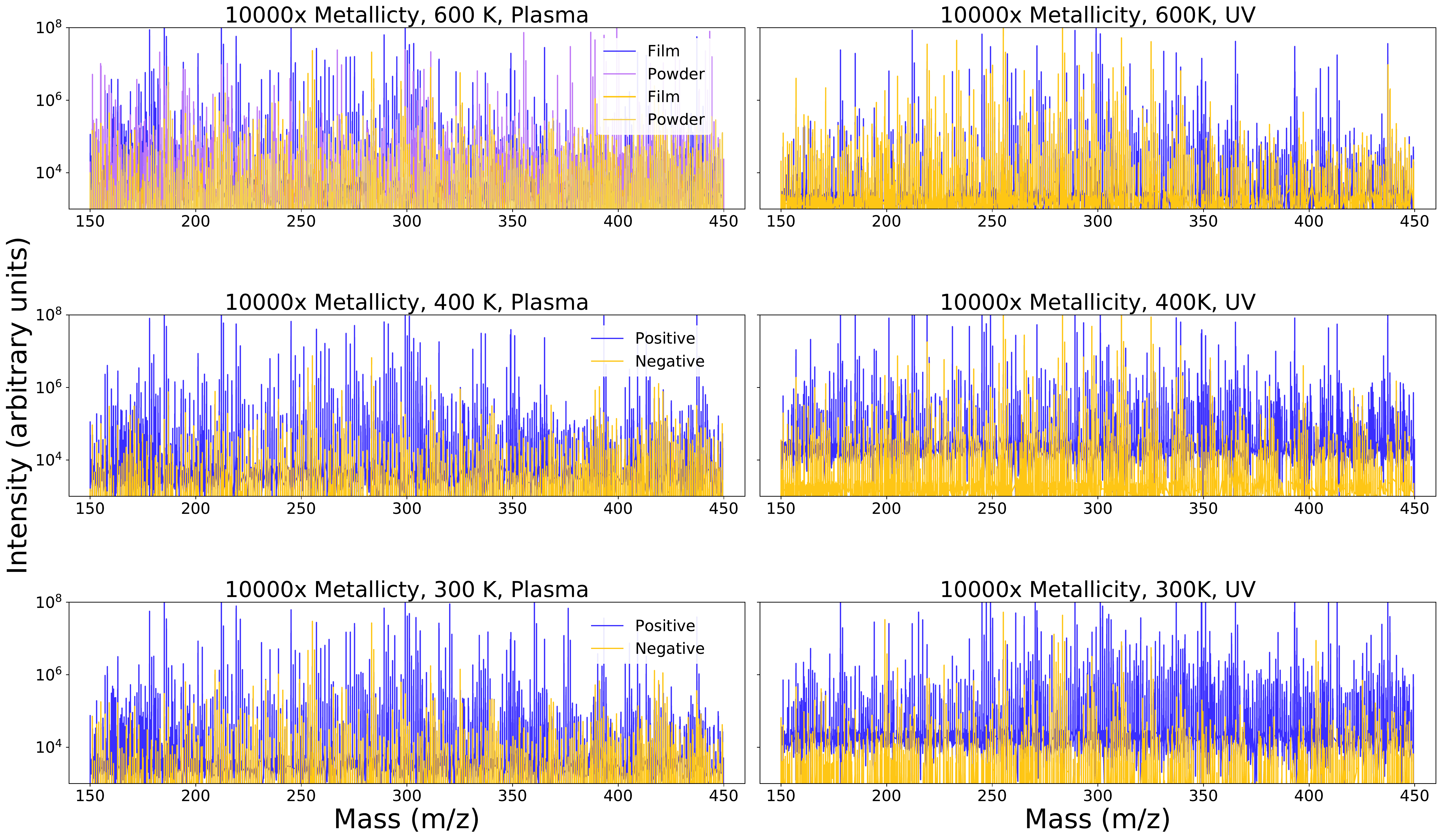}
\caption{Mass spectra from 150 to 450 m/z for all 10000$\times$ metallicity plasma and UV produced samples, dissolved in methanol. Both the 300 K and 400 K samples were determined to be soluble based on their mass spectra.}
\label{fig:ms_10000x_grid}
\end{figure*}

For the carbon dioxide-rich (10000$\times$ solar metallicity) inital gas mixtures, only the two lower temperature samples produced structured mass spectra, indicating soluble complex molecular compounds in the solid products. We observe repeating mass peak groups, though far less regular than that of the lower metallicity cases, of 10 u to 14 u in both the positive and negative ions of the data. The less regular mass peak groupings likely result from a weaker overall signal from these samples. Figure \ref{fig:ms_10000x_grid} shows mass spectra, while Table \ref{table:molecules_obscarbon} lists molecular formulas.

Carbon dioxide-rich cases produced very small amounts of sample, relative to the water-rich cases, and analysis was feasible for only two cases (300 K and 400 K). Nevertheless, in both samples we still find a number of prebiotic molecular formulas, including those for both derivatives of biological amino acids as well as non-proteinogenic amino acids. Notably, both soluble 10000$\times$ metallicity samples contain the formula for sugar alcohol glucitol, while the 300~K sample also contains the formula for glucose, the most common monosaccharide on Earth.

\begin{table*}[htbp]
\resizebox{\columnwidth}{!}{%
\begin{tabular}{ccccc}
\hline
\multicolumn{1}{l}{} & \multicolumn{1}{l}{} & \multicolumn{1}{l}{} & 
\multicolumn{1}{l}{10000$\times$ Results} & \multicolumn{1}{l}{} \\ 
\hline\hline

\multicolumn{1}{l}{} & \multicolumn{1}{l}{} & \multicolumn{1}{l}{} & \multicolumn{1}{l}{600 K \textit{material insoluble}} & \multicolumn{1}{l}{} \\

\hline\hline

\multicolumn{1}{l}{} & \multicolumn{1}{l}{} & \multicolumn{1}{l}{} & \multicolumn{1}{l}{400 K } & \multicolumn{1}{l}{} \\
\begin{tabular}[c]{@{}l@{}}
\textit{Mass (m/z) $\pm\Delta ppm$}\\
182.0790 $\pm0.1$ \\
192.1110 $\pm1.7$ \\
195.0895 $\pm0.1$\\
211.0845 $\pm0.1$\\
246.1216 $\pm1.2$\\
246.1328 $\pm1.2$ \\
267.1219 $\pm1.1$\\
276.1321 $\pm1.1$\\
\end{tabular} &
\begin{tabular}[c]{@{}l@{}}
\textit{Detection}\\
-/n \\
p \\
-\\
p \\
p \\
p \\
p \\
p \\
\end{tabular} &
\begin{tabular}[c]{@{}l@{}}
\textit{Formula}\\
C$_6$O$_6$H$_{14}$\\
C$_7$N$_2$O$_4$H$_{16}$\\
C$_{10}$NO$_3$H$_{13}$\\
C$_{10}$NO$_4$H$_{13}$\\
C$_{10}$N$_2$O$_5$H$_{18}$\\
C$_{9}$N$_4$O$_4$H$_{18}$\\
C$_{12}$N$_3$O$_4$H$_{17}$\\
C$_{11}$N$_2$O$_6$H$_{20}$\\
\end{tabular} & \begin{tabular}[c]{@{}l@{}}
\textit{Potential Molecule}\\
glucitol\\
orthinine acetate\\
tyrosine, O-methyl\\
tyrosine, 3-methoxy\\
Boc-L-glutamine\\
octopine\\
agaritine\\
saccharopine\\
\end{tabular} & \begin{tabular}[c]{@{}l@{}}
\textit{Relevance}\\
sugar alcohol\\
non-proteinogenic amino acid\\
non-proteinogenic amino acid\\
non-proteinogenic amino acid\\ 
non-proteinogenic amino acid\\
biological amino acid derivative\\
non-proteinogenic amino acid\\
biological amino acid derivative\\
\end{tabular}\\
\hline \hline 

\hline \hline

\multicolumn{1}{l}{} & \multicolumn{1}{l}{} & \multicolumn{1}{l}{} & \multicolumn{1}{l}{300 K } & \multicolumn{1}{l}{} \\
\begin{tabular}[c]{@{}l@{}}
\textit{Mass (m/z) $\pm\Delta ppm$}\\
180.0634 $\pm0.7$ \\
182.0790 $\pm0.2$ \\
192.1110 $\pm1.7$ \\
195.0895 $\pm0.1$ \\
246.1328 $\pm0.3$ \\
276.1321 $\pm1.1$ \\
\end{tabular} &
\begin{tabular}[c]{@{}l@{}}
\textit{Detection}\\
n \\
+/-\\
p\\
-\\
p\\
p\\
\end{tabular} &

\begin{tabular}[c]{@{}l@{}}
\textit{Formula}\\
C$_6$O$_6$H$_{12}$\\
C$_6$O$_6$H$_{14}$\\
C$_7$N$_2$O$_4$H$_{16}$\\
C$_{10}$NO$_3$H$_{13}$\\
C$_{9}$N$_4$O$_4$H$_{18}$\\
C$_{11}$N$_2$O$_6$H$_{20}$\\
\end{tabular} & \begin{tabular}[c]{@{}l@{}}
\textit{Potential Molecule}\\
glucose\\
glucitol\\
orthinine acetate\\
tyrosine, O-methyl\\
octopine\\
saccharopine\\
\end{tabular} & \begin{tabular}[c]{@{}l@{}}
\textit{Relevance}\\
monosaccharide\\
sugar alcohol\\
non-proteinogenic amino acid\\
non-proteinogenic amino acid\\
biological amino acid derivative\\
biological amino acid derivative\\
\end{tabular}\\
\hline \hline
\end{tabular}}
\caption{Molecular formulas detected from the 10000$\times$ metallicity experiments.}
\tablenotetext{}{Detection column indicates energy source and detection polarity. Plasma (+: positive ion, -: negative ion) and UV (p: positive ion, n: negative ion). We report the smaller $\Delta$ppm between measured m/z and exact m/z when a detection was made in more than one instrument mode.}
\label{table:molecules_obscarbon}
\end{table*}
\newpage
\section{Discussion}
\subsection{Solubility of Exoplanet Haze Analogues} \label{subsection:solubility}

As shown in Figure \ref{fig:solubility}, the exoplanet haze analogues produced in the laboratory with the plasma energy source exhibited diverse solubility behavior. Quantitative measurements of solubility were outside the scope of this study. Additionally, there are various inconsistencies in the literature about protocols used to determine solubility of the complex mixtures often referred to as  ``tholin'' \citep[for discussion see e.g.,][]{carrasco2009solubility, he2014solubilitytholin}. Instead our solubility metric was qualitative, determined both by visual inspection of the sample within the solvent as well as visual inspection of the resulting mass spectral data (which can be affected by other chemical properties). During visual inspection of a sample in a solvent, we noted any color change and any visible decrease in the amount of solid. We also visually compared the mass spectrum for each sample in solution with corresponding results for a control blank solvent (see Figure \ref{fig:blank_ms}). From these post-measurement observations, we made a determination about the fidelity of the signal, and thus whether any sample had dissolved in the solvent. As the UV energy source produces significantly less sample \citep{he2018hazeuvplasma}, solubility observations of the kind performed here were not possible.


As discussed in detail below, the lower temperature plasma samples always appeared soluble while the higher temperature samples were more likely to resist dissolving in a particular solvent. The hydrogen-dominant (100$\times$ solar metallicity) initial gas mixtures yielded methanol-insoluble solid haze particles except for the 300 K condition, which notably includes trace amounts of NH$_3$, ammonia. For the water-rich (1000$\times$ metallicity) cases, all solid samples appeared soluble in methanol, the first choice solvent for measurements. This solubility behavior  is similar to Titan-like ``tholin'' haze analogues that result from nitrogen gas mixtures with trace amounts of methane and carbon monoxide \citep{carrasco2009solubility}, which often demonstrate significant solubility in methanol. Finally, the two lower temperature cases for the CO$_2$-dominant (10000$\times$ metallicity) gas mixtures were somewhat soluble in methanol, while the highest temperature 600 K condition yielded highly insoluble solid products. 


Both polar solvents and polar-nonpolar mixtures were tested. Samples were only soluble in the pure polar solvents. From previous measurements of the particle structure \citep{he2018particle}, long chains of particles were observed for the water-rich (1000$\times$) 300 K and 400 K solid products. This structure suggests that the compounds themselves are polar, and thus their high solubility in the polar solvent of methanol is consistent with the general chemical rule ``like dissolves like''. The other polar solvent, dichloromethane, was also effective at dissolving the solid haze analogue samples. Titan ``tholin'' also exhibits highly polar solubility \citep[see e.g.,][]{carrasco2009solubility}, marking an additional similarity in the broad chemical behavior of our exoplanet ``tholin''. While the 1000$\times$ metallicity exoplanet analogues share some physical characteristics with Titan haze analogues, elemental analysis (see Table \ref{table:elementalratiosplasma}) shows they are chemically distinguishable.

While earlier works suggested that the soluble fraction of ``tholin'' from Titan and similar planetary atmospheric experiments were representative of the sample as a whole \citep{carrasco2009solubility}, more recent studies found that the soluble and insoluble fractions may be chemically distinct \citep{somogyi2016,maillard2018soluble}. This suggests that in addition to the limitations of our study regarding the solubility of our samples, the data we do have may not reveal the full chemical complexity of our exoplanet haze analogues. 
Future work on the chemistry of exoplanetary hazes should consider additional measurements that are not solubility dependent. For example, laser desorption/ionization (LDI) mass spectrometry measurements do not require soluble sample and have successfully identified insoluble macromolecules in Martian meteorite samples \citep{somogyi2016}.

In addition to practical experimental considerations, the solubility of planetary haze analogues has further implications for planetary atmospheres themselves. For example, haze particles are known to act as cloud condensation nuclei (CCN) in many atmospheres, such as the organic haze for ethane/methane clouds on Titan \citep{horsttitanreview}, meteoritic smoke particles for water ice clouds on Mars \citep{hartwickmarsccn2019}, and sand storms and seaspray for low lying clouds on Earth \citep{hellingreview2019}. Solids that are soluble in the atmospheric condensates of a world (such as salt in seaspray in water vapor on Earth) promote cloud formation and enable the creation of significant cloud belts. These condensation seeds facilitate cloud formation by reducing the level of saturation required for cloud materials to condense \citep{hellingreview2019}. The production of polar soluble solid haze particles high in the atmosphere, as analogous to the experiments considered here, may suggest that polar condensible atmospheric constituents may more easily form clouds in exoplanet environments similar to our experimental atmospheres. For example, the hazes produced in our laboratory simulations might promote water cloud formation in cool enough atmospheres, which would be particularly relevant to our 300 K temperature regime across all metallicity conditions. Both the composition of the insoluble experimental hazes and their effectiveness as cloud seed particles are avenues for future study.

\subsection{Prebiotic Material in Exoplanet Haze Analogues} \label{subsection:prebiotic}
Some of the first investigations of prebiotic chemistry assumed that synthesis required liquid water to occur \citep{miller1953,millerurey1959}. However, aerosols have long since been recognized as a source of prebiotic material, including amino acids, nucleobases, sugars, purines, and pyrimidines on the early Earth \citep[e.g.,][]{Dobsonprebioticaerosol2000}. Mass spectrometry has been used in a variety of exobiology focused investigations from meterorites to Mars to Titan \citep[e.g.,][and references therein]{Sarkertitanmassspec,Neish2010massspectitanhydrolysis, CallahanMassSpecMeteorites,vuitton2014_massspec_inspace,somogyi2016}, and its use has successfully enabled identification of amino acids and nucleobases in the products of Titan atmosphere simulation experiments \citep{horst2012formation,sebree2018}, as well as both amino acids and sugars in meteorite samples suggested to have seeded the early Earth \citep{cooper2001sugarearlyearth}.

Here, we have identified molecular formulas for eight biological amino acids (tyrosine, tryptophan, histidine, pyrrolysine, lysine, arginine, glutamine, glutamic acid) as well as dozens of their derivatives and two nucleobase formulas (guanine and adenine) as well as the formula for a derivative (thymine glycol deriving from thymine). We also detect, for the first time in the products of an atmospheric experiment that did not contain liquid water, the molecular formulas for simple sugar molecules and sugar derivatives (collectively called polyols): glyceraldehyde, gluconic acid, sucrose, glucitol, and glucose. 

Previous laboratory simulations of UV-radiated precometary ice analogues \citep{meinert2016riboseice,demarcellus2015sugarice,nuevo2018sugarice} and laboratory simulations of high velocity impacts \citep[e.g.,][]{civis2016sugarsimpacts,ferus2019prebioticformaldehydeimpacts} have detected numerous saccharides including ribose and deoxyribose, bolstering the theory that prebiotic planetary chemistry relies on external delivery via cometary, meteoritic, or interplanetary dust sources. Moreover, analysis of extraterrestrial sources such as the Murchison and Murray meteorites also shows the presence of both simple sugars, sugar alcohols, and sugar acids \citep{cooper2001sugarearlyearth,cooper2018sugarreview}. Even more recently, bioessential sugars such as ribose and other pentoses have also been found in Murchison and NWA 801 meteorite samples \citep{Furukawa2019sugarmurchison}. Other probes of external delivery sources farther afield than the local solar system neighborhood also exist. The simplest sugar-related molecule, glycolaldehyde, has been detected in interstellar molecular clouds \citep{hollis2000sugarism}, and amines and amides have been detected throughout the interstellar medium \citep{sun2016organicsreview}. In addition, the well-studied formose reaction, in which formaldehyde reacts to form a multitude of sugar molecules, has been studied both for interstellar synthesis of sugars in the gas phase \citep{jalbout2007formosegasphase} as well as extensively in aqueous solutions mimicking hydrothermal vents deep in the prehistoric Earth's ocean \citep{kopetzki2011hydrothermalformose} and under more temperate alkaline liquid water conditions \citep{Pestunova2005formoseUVrad}.

Our results suggest that, given the right mixture of gases, a planetary atmosphere alone could photochemically generate not only amino acids and nucleobases, but even simple sugars. While not discounting external delivery of prebiotic materials, this result underscores the idea that at least preliminary abiogenesis can occur both in interstellar space and via external delivery as well as \textit{in situ} in the atmospheres of planets themselves. The yields of any such prebiotic materials made in planetary atmospheres would require careful consideration, however \citep[e.g.,][]{harman2013glycoinearlyearth}, and further reactions to generate more complex sugars and eventually biomolecules still would likely require liquid water and remain challenging \citep{schwartz2007yields}.

Our 300 K, 100$\times$ metallicity simulated atmosphere, which is primarily hydrogen with lesser amounts of water and methane and trace amounts of ammonia, produced the formula for the simplest monosaccharide, glyceraldehyde (C$_3$H$_6$O$_3$). This atmosphere, with its high H$_2$ content, is likely most analogous to that of a mini-Neptune. The heavier metallicity experimental atmospheres (likely more analogous to super-Earth atmospheres) containing larger amounts of water, carbon dioxide, and methane were additionally able to produce more complex sugar molecular formulas such as glucose, sucrose, glucitol, and gluconic acid. Notably, these molecular formulas all occur across our range of temperatures from 600 K to 300 K. 

We advise a note of caution in all our reported molecular detections. We detect only the formulas for all of the molecules in Tables \ref{table:molecules_obshydro}, \ref{table:molecules_obswater1}, \ref{table:molecules_obswater2}, \ref{table:molecules_obswater3}, \ref{table:molecules_obswater4}, and \ref{table:molecules_obscarbon}. Our instrumental set-up alone cannot confirm molecular structure. High resolution mass spectrometry, as performed here, gives very precise molecular mass measurements. However, with complex mixtures of the kind examined here, many possible molecular combinations exist and overlap. Identifications that rely on mass only for such complex mixtures are therefore highly degenerate. Verification of the prebiotic molecules discussed here will involve follow-up measurements with other techniques that can infer and isolate molecular structure, such as high performance liquid chromatography (HPLC). 
\vspace{10pt}

\subsection{Chemical Pathways to Haze Formation}

Gas phase results from these laboratory experiments have already been published \citep{he2019gasphase}, and allow us to hypothesize some chemical pathways for the formation of the solid hazes discussed here. Our gas phase study found that abiotic production of oxygen, organics, and prebiotic molecules occurs readily in these mini-Neptune to super-Earth analogue atmospheres, suggesting that even the co-presence of such molecules ought not to be taken as a biosignature. The results presented here about the ability of such atmospheres to form sugars, amino acids, and nucleobases shows that while the presence of such ``false positive'' biosignature gases should be treated with skepticism, they also allow for the formation of a rich prebiotic inventory. The remaining steps from prebiotic chemistry to biology remains an open question. Observers in future exoplanet studies must balance biosignature searches with the knowledge that while abiotic production must always first be ruled out, the coexistence of such gases may also indicate that prebiotic chemistry has progressed significantly in the atmosphere and could further develop on any putative surface.

For the conditions in which we detected the formula for glyceraldehyde (the 300 K 100$\times$ and 1000$\times$ experiments), the gas phase results showed increased production of hydrogen cyanide (HCN) and formaldehyde (HCHO) \citep{he2019gasphase}, both known to participate in the generation of sugars \citep{schwartz1984hcnprebiotic, cleaves2008prebioticformaldehyde}. The production of the variety of amino groups in the solid phase is also unsurprising given the number of organic precursors observed in the gas phase.

\begin{figure*}[ht!]
\centering
\includegraphics[angle=0,width=0.9\linewidth]{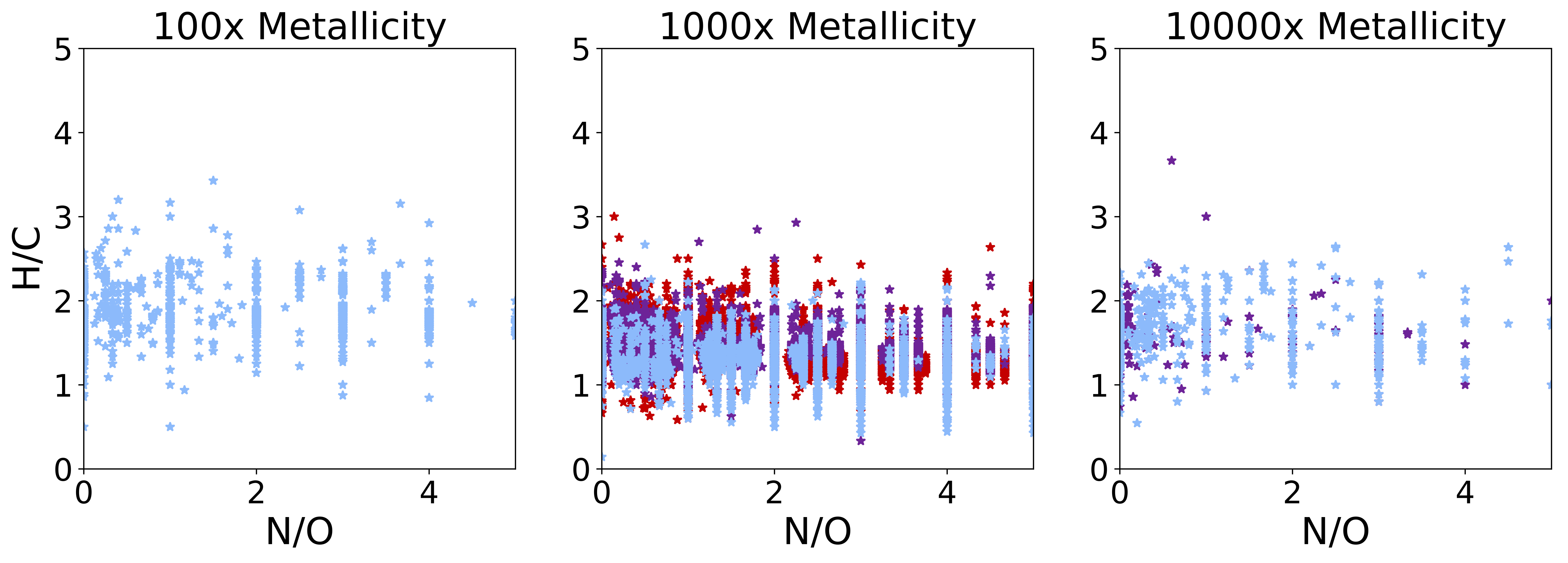}
\includegraphics[angle=0,width=0.99\linewidth]{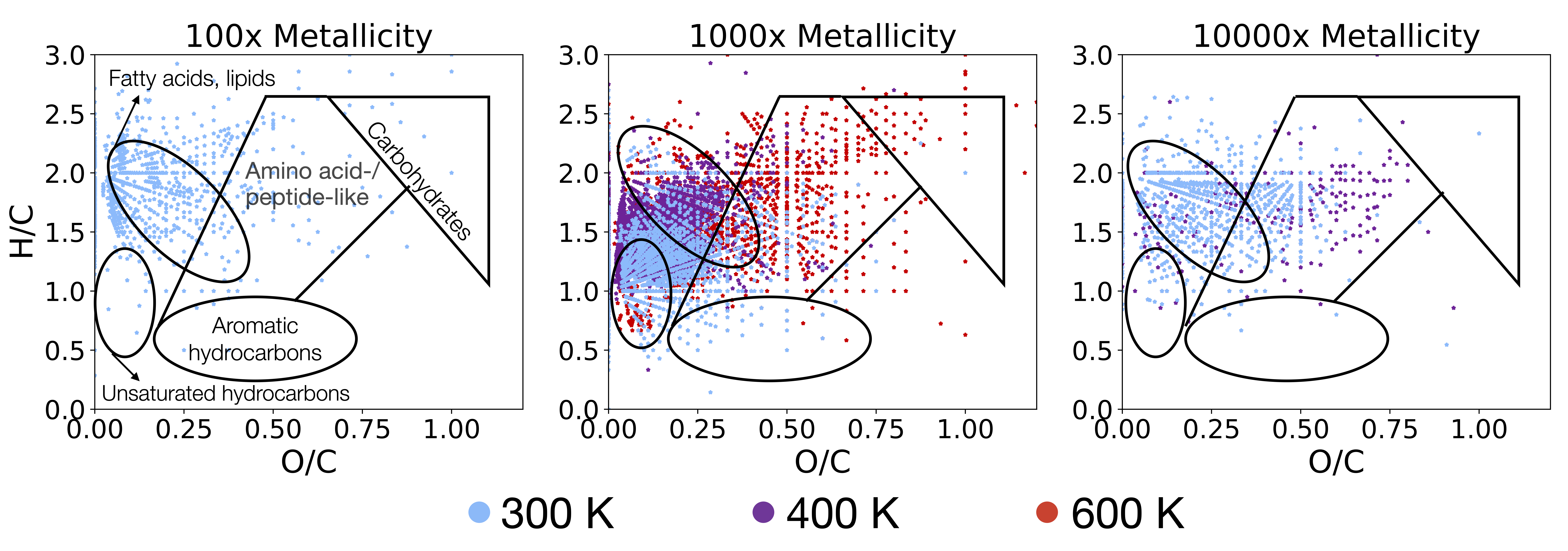}
\caption{Van Krevelen diagrams of each measured set of samples, showing the hydrogen-to-carbon vs nitrogen-to-oxygen ratios (top row) and hydrogen-to-carbon vs oxygen-to-carbon ratios (bottom row) in each set of solid haze analogue material. Red symbols correspond to the 600 K samples, purple to 400 K samples, and blue to 300 K samples. The labels of compound regions on the lower 100$\times$ plot apply to the entire lower row.}
\label{fig:vk}
\end{figure*}

Between different atmospheres, we observe that the solid haze analogues appear to incorporate certain molecules more readily than others. In Figure \ref{fig:vk}, we show Van Krevelen diagrams, which are widely used in the petroleonomics field and have since been used for Titan atmospheric haze studies \citep[e.g.,][]{horstthesis}. These diagrams help visualize classes of compounds, as these have characteristic elemental ratios, resulting in clustering of similar compounds in specific locations on a Van Krevelen diagram \citep{Kim2003VKoverview}. We show two forms of this diagram. The first (top row of Figure \ref{fig:vk}) compares H/C to N/O, which shows that distinct nitrogen-to-oxygen ratios form in all of our samples. The second (bottom row of Figure \ref{fig:vk}) shows H/C vs O/C. We follow the ratio bounds of \citet{Ruf2018VKastro} to map where carboxylic acids (fatty acids), unsaturated hydrocarbons, aromatic hydrocarbons, amino-acid-like compounds, and carbohydrates/sugars fall in this H/C vs O/C phase space. Distinct diagonal, vertical, and horizontal lines are visible in this phase space as well. Such lines form along characteristic H/C and O/C ratios, which are characteristic of particular reaction pathways, such as methylation or demethylation, oxidation or reduction, etc. \citep{Kim2003VKoverview}.

Interestingly, in the 1000$\times$ experiments, in the initial gas mixtures oxygen increases with decreasing temperature, but the opposite is seen in the elemental analysis of the solid haze products. From the least oxygen-rich and most carbon-rich gas mixture (the 600 K case), we see the strongest incorporation of oxygen into the solid, while with the least carbon-rich and most oxygen-rich gas mixture (the 300 K case), we see the least oxygen. This is clearly observed in the middle panel of the bottom row of Figure \ref{fig:vk}, which is in part why we are able to identify so many amino-acid-like formulas in this sample. Previous experiments on haze formation found that the increasing presence of carbon monoxide promotes aerosol production \citep{horsttolbert2014}, which the authors speculate could occur by shifting the oxygen incorporation more readily into the solid phase. Notably, the 1000$\times$ experiment showing the largest oxygen solid content (the 600 K case) is the only initial gas mixture to contain carbon monoxide, which is consistent with this interpretation by \citet{horsttolbert2014}. This further suggests that not only does the initial gas mixture matter in terms of the elemental species present, but the molecular carriers of these species matters greatly as well because these molecular carriers determine which elements are able to participate effectively in haze formation.

Furthermore, the role of nitrogen in haze formation is clearly very important, yet poorly understood \citep[e.g.,][]{Imanaka2007,trainer2012nitrogen,horst2018titanlabparticlegas}. Our only 100$\times$ experiment to produce soluble haze products is also the only 100$\times$ experiment that contained a nitrogen-bearing molecule in the initial gas mixture, NH$_3$. With the UV energy source, this nitrogen-containing gas mixture also had the highest production rate of the 100$\times$ conditions and the second highest production rate of any condition \citep{he2018hazeuvplasma}. This 300 K, 100$\times$ experiment produced the formulas for a nucleotide base, a monosaccharide, and both biological and non-proteinogenic amino acids, underscoring the dramatic role nitrogen can play in haze formation. Its ability to change the solubility of the hazes produced may have additional implications for its role in the chemistry of the system as well.

In comparison to previous Titan work, all our haze samples across metallicities and temperatures have far more oxygen, as shown in Figure \ref{fig:elemental_analysis}. Although some of the physical characteristics of the haze products are similar, i.e., production rate \citep{horst2018production}, color and particle size \citep{he2018particle}, and solubility (this work), our elemental analysis demonstrates robustly that these haze analogues are quite distinct chemically. Therefore, modeling efforts that use so-called ``hydrocarbon'' haze as a proxy in exoplanet studies must practice due caution as the optical properties and spectroscopic impact of true exoplanet hazes, at least for a wide range of super-Earths and mini-Neptunes, will likely also be different than that of hydrocarbon hazes. Furthermore, aside from incorrect observational interpretations from exoplanet transmission, emission, or reflectance spectroscopy, the chemical interpretations of such worlds will also be misconstrued if hydrocarbon haze proxies are used. Finally, such photochemical modeling efforts typically only include up to C$_6$ species due to computational complexity and expense as well as a lack of required information such as reaction rates and photolysis cross sections \citep{vuitton2019}. However, recent work exploring Titan-like hazes shows that heavy molecular weight compounds ($\geq$ C$_8$) are needed to fully explain aerosol formation and growth for Titan's haze \citep{berry2019}. Considering the significant addition of oxygen in the exoplanet simulations performed here, inclusion of heavy molecular weight compounds is likely paramount to properly capturing exoplanet hazes through photochemical modeling.


\subsection{Influence of Different Energy Sources on Haze Formation and Composition}

Our results show general broad agreement within error between the elemental composition of hazes produced with the AC plasma and the FUV lamp. As noted previously, the AC plasma is not directly mimicking any specific process, but is instead a proxy for highly energetic upper planetary atmospheres. This suggests that the overall elemental composition of atmospheric hazes may not be as strongly affected by the source of energy imparted onto the atmosphere. However, the specific molecular identifications we are able to make does vary greatly between the two energy sources. As the energy density of the UV lamp is lower than that of the plasma \citep{he2019gasphase}, the plasma source typically produces both more and larger solid particles \citep{he2018particle, he2018hazeuvplasma}. This may contribute to our seeing many more molecules in the mass spectral data of the plasma produced particles, as well as our ability to make molecular identifications. Additionally, because there is less sample to dissolve in the solvent, the concentration of the UV produced haze analogues injected into the mass spectrometer is typically lower than that of the plasma. These results are consistent with a study of Titan-like aerosols, in which the use of UV-photolysis as the energy source also generated fewer MS/MS detected prebiotic molecules than did plasma-produced aerosols \citep{sebree2018}. Because the UV light source imparts less energy into the system, longer experimental steady states (beyond what is practical for the laboratory) would likely be required to generate more complex molecules. In a real planetary atmosphere, such timescales of UV photon bombardment may be less of an issue. However, the gases in the atmosphere will still require sufficiently energetic UV fluxes to dissociate their bonds to produce photochemical aerosols of any complexity.

These results may have implications for the ability of various stellar types of stars to induce complex photochemistry on their hosted planets. Recent modeling of quiescent M-dwarfs with less intense UV fluxes has shown that reaction rates for prebiotic chemistry on planets around M-dwarfs should be slower and that prebiotic pathways may in some cases be unable to proceed at all \citep{ranjan2017uvmdwarf}. Additional studies comparing the reaction rates of known pyrimidine synthesis in the presence and absence of UV light reiterates this result for M-dwarf planets, though they also consider whether the more frequent powerful flaring events on M-dwarfs may be enough to overcome this lack of quiescent UV flux during most of the stellar lifetime \citep{rimmer2018rna}. 

Another complication is that currently, we have relatively few measurements of planet host spectra in the UV. While these data gaps can be overcome with modeling approaches \citep[e.g.,][]{peacock2019uvradiation} and additional observational campaigns \citep[e.g.,][]{youngblood2017uvflaresmdwarfs}, translating these UV fluxes into proxies usable in the laboratory remains challenging. Any close-in exoplanet would likely be subject to charged particles traveling along stellar magnetosphere lines or bombardment by cosmic rays. These high energy particles could induce prolific chemistry in a planetary atmosphere. However, constraining the rates and magnitudes of such energetic particles deposited into the atmosphere is also outside the ability of current observations, and thus quantifying this energy flux for use in laboratory simulations is also difficult.

\subsection{Prospects for the Observability of Exoplanet Haze Chemistry}
As shown in this work, we expect a broad range of hazes over the diverse phase space of exoplanet atmospheres. While the chemistry described here is intriguing for exoplanet studies, there is currently a disconnect between laboratory production of these haze analogues and detection of these materials in exoplanet observations. Future measurements to obtain the optical properties of these hazes will provide observers with spectral features to search for with future spectroscopic observatories such as the \textit{James Webb Space Telescope (JWST)}, the \textit{ARIEL Space Telescope}, or the \textit{Wide Field Infrared Survey Telescope (WFIRST)} beyond merely the muting of spectral features as observed thus far \citep[e.g.,][]{kreidberg2014clouds}. Moreover, such optical property measurements would provide an additional layer of confirmation as to the presence of various chemical bonds in the haze particles and would thus provide additional evidence for our compositional measurements performed here. 

The ubiquity of planetary hazes will impact both transiting exoplanet studies as well as future direct imaging missions to obtain spectra of exoplanet atmospheres in reflected light. Our experiments show substantial differences in haze production \citep{horst2018production,he2018hazeuvplasma}, likely leading to impacts for observations across a diverse range of atmospheres. However,
observations of the atmospheres of mini-Neptunes and super-Earths to obtain their gas composition across wide wavelength ranges that probe different pressures in the atmosphere may also help reveal whether any substantial photochemistry is occurring on the planet. As we now have both gas phase chemistry \citep{he2019gasphase} and solid phase chemistry constraints (this work) for photochemistry of a subset of these atmospheres, we may begin to infer the presence or absence of these processes from observations. 

Observations of mini-Neptune atmospheres have just begun to produce compositional constraints. For example, a recent study of GJ~3470~b found a hydrogen-dominated atmosphere with depleted water, ammonia, and methane gas and Mie scattering aerosols \citep{benneke2019h2subneptune}, reminiscent of the gas phase chemistry of our 300 K, 100$\times$ metallicity experiment \citep{he2019gasphase}, though this planet has a much higher equilibrium temperature of nearly 700 K. Another cooler ($\sim$300 K) mini-Neptune, K2-18  b, was recently shown to have significant water and possible water clouds in its atmosphere \citep{benneke2019k2,tsiaras2019k2}, showing the diversity of mini-Neptune atmospheres. 

This diversity may result from temperature differences between the planets resulting in differing atmospheric chemistry, as shown is likely from our laboratory experiments. When optical properties of the hazes discussed here are obtained, these exoplanets would make fascinating targets for future observatories. With observations of both individual planets as case studies and of larger planetary trends in temperature and atmospheric composition, we can explore whether any of the hazes we find experimentally are truly present in existent exoplanetary atmospheres and thus further investigate the prevalence of various chemical pathways. 
\newpage
\section{Conclusion} \label{sec:conclusions}
We have conducted very high resolution Orbitrap mass spectrometry measurements of the solid haze products resulting from a suite of laboratory experiments from the PHAZER chamber, exploring temperate exoplanet atmospheres over a range of initial gas chemistries. We find that these haze products show varying solubility behavior, with all solids being at least partially soluble in polar solvents, suggesting that these hazes may make for effective cloud condensation nuclei in exoplanetary atmospheres with polar condensible material. Additionally, we find that all haze products have very large oxygen contents in the solid products, showing a marked difference in elemental composition to previous Titan atmospheric work. Finally, we detect a number of prebiotic molecular formulas, including those for biological and non-proteinogenic amino acids, for two nucleotide bases, and for the first time from an atmospheric experiment without liquid water, formulas for simple sugars. 

This work demonstrates the power of laboratory experiments in understanding the complex chemistry at work in exoplanet atmospheres, both at large general scales as well as at for detailed single compound detections. Future follow up work is required to confirm the presence of our prebiotic molecular formula detections, as well as to understand the ability of haze particles to act as cloud condensation nuclei in such atmospheres. Connecting the chemical information gathered here to a telescopic observable will be highly important to make the most of these results and their implications for distant worlds.

\acknowledgments
S.E. Moran was supported by NASA Earth and Space Science Fellowship Grant 80NSSC18K1109. Portions of this study were supported by NASA Exoplanets Research Program Grant NNX16AB45G. C.~He was supported by the Morton K. and Jane Blaustein Foundation. This work is supported by the French National Research Agency in the framework of the Investissements d'Avenir program (ANR-15-IDEX-02), through the funding of the ``Origin of Life'' project of the Univ. Grenoble-Alpes and the French Space Agency (CNES) under their Exobiology and Solar System programs. C{\'e}dric Wolters acknowledges a PhD fellowship from CNES/ANR (ANR-16-CE29-0015 2016-2021) S.E. Moran also thanks the entire H{\"o}rst PHAZER lab group and the STARGATE collaboration for their useful discussion and support on this work, as well as two anonymous referees whose careful review improved this manuscript.

\bibliography{apjmnemonic,orbi}

\begin{thebibliography}{}
\expandafter\ifx\csname natexlab\endcsname\relax\def\natexlab#1{#1}\fi

\bibitem[{{Arney} {et~al.}(2017){Arney}, {Meadows}, {Domagal-Goldman},
  {Deming}, {Robinson}, {Tovar}, {Wolf}, \&
  {Schwieterman}}]{arney2017paleorange}
{Arney}, G.~N., {Meadows}, V.~S., {Domagal-Goldman}, S.~D., {et~al.} 2017,
  \apj, 836, 49

\bibitem[{{Benneke} {et~al.}(2019{\natexlab{a}}){Benneke}, {Knutson},
  {Lothringer}, {Crossfield}, {Moses}, {Morley}, {Kreidberg}, {Fulton},
  {Dragomir}, \& {Howard}}]{benneke2019h2subneptune}
{Benneke}, B., {Knutson}, H.~A., {Lothringer}, J., {et~al.} 2019{\natexlab{a}},
  Nature Astronomy, 361

\bibitem[{{Benneke} {et~al.}(2019{\natexlab{b}}){Benneke}, {Wong}, {Piaulet},
  {Knutson}, {Lothringer}, {Morley}, {Crossfield}, {Gao}, {Greene}, {Dressing},
  {Dragomir}, {Howard}, {McCullough}, {Kempton}, {Fortney}, \&
  {Fraine}}]{benneke2019k2}
{Benneke}, B., {Wong}, I., {Piaulet}, C., {et~al.} 2019{\natexlab{b}}, \apjl,
  887, L14

\bibitem[{{Berry} {et~al.}(2019){Berry}, {Ugelow}, {Tolbert}, \&
  {Browne}}]{berry2019}
{Berry}, J.~L., {Ugelow}, M.~S., {Tolbert}, M.~A., \& {Browne}, E.~C. 2019,
  \apjl, 885, L6

\bibitem[{{Bonnet} {et~al.}(2013){Bonnet}, {Thissen}, {Frisari}, {Vuitton},
  {Quirico}, {Orthous-Daunay}, {Dutuit}, {Le Roy}, {Fray}, {Cottin},
  {H{\"o}rst}, \& {Yelle}}]{bonnet2013compositional}
{Bonnet}, J.-Y., {Thissen}, R., {Frisari}, M., {et~al.} 2013, International
  Journal of Mass Spectrometry, 354, 193

\bibitem[{{Borucki} {et~al.}(2010){Borucki}, {Koch}, {Basri}, {Batalha},
  {Brown}, {Caldwell}, {Caldwell}, {Christensen-Dalsgaard}, {Cochran},
  {DeVore}, {Dunham}, {Dupree}, {Gautier}, {Geary}, {Gilliland}, {Gould},
  {Howell}, {Jenkins}, {Kondo}, {Latham}, {Marcy}, {Meibom}, {Kjeldsen},
  {Lissauer}, {Monet}, {Morrison}, {Sasselov}, {Tarter}, {Boss}, {Brownlee},
  {Owen}, {Buzasi}, {Charbonneau}, {Doyle}, {Fortney}, {Ford}, {Holman},
  {Seager}, {Steffen}, {Welsh}, {Rowe}, {Anderson}, {Buchhave}, {Ciardi},
  {Walkowicz}, {Sherry}, {Horch}, {Isaacson}, {Everett}, {Fischer}, {Torres},
  {Johnson}, {Endl}, {MacQueen}, {Bryson}, {Dotson}, {Haas}, {Kolodziejczak},
  {Van Cleve}, {Chandrasekaran}, {Twicken}, {Quintana}, {Clarke}, {Allen},
  {Li}, {Wu}, {Tenenbaum}, {Verner}, {Bruhweiler}, {Barnes}, \&
  {Prsa}}]{BoruckiKepler}
{Borucki}, W.~J., {Koch}, D., {Basri}, G., {et~al.} 2010, Science, 327, 977

\bibitem[{{Cable} {et~al.}(2012){Cable}, {H{\"o}rst}, {Hodyss}, {Beauchamp},
  {Smith}, \& {Willis}}]{cabletholinreview2012}
{Cable}, M.~L., {H{\"o}rst}, S.~M., {Hodyss}, R., {et~al.} 2012, Chemical
  Reviews, 3, 1882

\bibitem[{{Callahan} {et~al.}(2011){Callahan}, {Smith}, {Cleaves}, {Ruzicka},
  {Stern}, {Glavin}, {House}, \& {Dworkin}}]{CallahanMassSpecMeteorites}
{Callahan}, M.~P., {Smith}, K.~E., {Cleaves}, H.~J., {et~al.} 2011, Proceedings
  of the National Academy of Science, 108, 13995

\bibitem[{{Carrasco} {et~al.}(2009){Carrasco}, {Schmitz-Afonso}, {Bonnet},
  {Quirico}, {Thissen}, {Dutuit}, {Bagag}, {Lapr{\'e}vote}, {Buch}, {Giulani},
  {Adand {\'e}}, {Ouni}, {Hadamcik}, {Szopa}, \&
  {Cernogora}}]{carrasco2009solubility}
{Carrasco}, N., {Schmitz-Afonso}, I., {Bonnet}, J.~Y., {et~al.} 2009, Journal
  of Physical Chemistry A, 113, 11195

\bibitem[{{Civi{\v{s}}} {et~al.}(2016){Civi{\v{s}}}, {Szabla}, {Szyja},
  {Smykowski}, {Ivanek}, {Kn{\'\i}{\v{z}}ek}, {Kubel{\'\i}k}, {{\v{S}}poner},
  {Ferus}, \& {{\v{S}}poner}}]{civis2016sugarsimpacts}
{Civi{\v{s}}}, S., {Szabla}, R., {Szyja}, B.~M., {et~al.} 2016, Scientific
  Reports, 6, 23199

\bibitem[{{Cleaves}(2008)}]{cleaves2008prebioticformaldehyde}
{Cleaves}, H.~James, I. 2008, Precambrian Research, 164, 111

\bibitem[{{Cloutier} \& {Menou}(2019)}]{cloutierandmenou2019}
{Cloutier}, R., \& {Menou}, K. 2019, arXiv e-prints, arXiv:1912.02170

\bibitem[{{Cooper} {et~al.}(2001){Cooper}, {Kimmich}, {Belisle}, {Sarinana},
  {Brabham}, \& {Garrel}}]{cooper2001sugarearlyearth}
{Cooper}, G., {Kimmich}, N., {Belisle}, W., {et~al.} 2001, \nat, 414, 879

\bibitem[{Cooper {et~al.}(2018)Cooper, Rios, \& Nuevo}]{cooper2018sugarreview}
Cooper, G., Rios, A.~C., \& Nuevo, M. 2018, Life, 8, doi:10.3390/life8030036

\bibitem[{{de Marcellus} {et~al.}(2015){de Marcellus}, {Meinert},
  {Myrgorodska}, {Nahon}, {Buhse}, {d'Hendecourt}, \&
  {Meierhenrich}}]{demarcellus2015sugarice}
{de Marcellus}, P., {Meinert}, C., {Myrgorodska}, I., {et~al.} 2015,
  Proceedings of the National Academy of Science, 112, 965

\bibitem[{{Demory} {et~al.}(2016){Demory}, {Gillon}, {de Wit}, {Madhusudhan},
  {Bolmont}, {Heng}, {Kataria}, {Lewis}, {Hu}, {Krick}, {Stamenkovi{\'c}},
  {Benneke}, {Kane}, \& {Queloz}}]{demory2016}
{Demory}, B.-O., {Gillon}, M., {de Wit}, J., {et~al.} 2016, \nat, 532, 207

\bibitem[{{Dobson} {et~al.}(2000){Dobson}, {Ellison}, {Tuck}, \&
  {Vaida}}]{Dobsonprebioticaerosol2000}
{Dobson}, C.~M., {Ellison}, G.~B., {Tuck}, A.~F., \& {Vaida}, V. 2000,
  Proceedings of the National Academy of Science, 97, 11864

\bibitem[{{Dragomir} {et~al.}(2015){Dragomir}, {Benneke}, {Pearson},
  {Crossfield}, {Eastman}, {Barman}, \& {Biddle}}]{dragomir2015rayleigh}
{Dragomir}, D., {Benneke}, B., {Pearson}, K.~A., {et~al.} 2015, Astrophysical
  Journal, 814, 102

\bibitem[{{Elkins-Tanton} \& {Seager}(2008)}]{ElkinsTanton2008super-earths}
{Elkins-Tanton}, L.~T., \& {Seager}, S. 2008, \apj, 685, 1237

\bibitem[{{Ferus} {et~al.}(2019){Ferus}, {Pietrucci}, {Saitta}, {Ivanek},
  {Knizek}, {Kubel{\'\i}k}, {Krus}, {Juha}, {Dudzak}, {Dost{\'a}l}, {Pastorek},
  {Petera}, {Hrncirova}, {Saeidfirozeh}, {Shestivsk{\'a}}, {Sponer}, {Sponer},
  {Rimmer}, {Civi{\v{s}}}, \&
  {Cassone}}]{ferus2019prebioticformaldehydeimpacts}
{Ferus}, M., {Pietrucci}, F., {Saitta}, A.~M., {et~al.} 2019, \aap, 626, A52

\bibitem[{{Fleury} {et~al.}(2019){Fleury}, {Gudipati}, {Henderson}, \&
  {Swain}}]{fleury2019HJexperiments}
{Fleury}, B., {Gudipati}, M.~S., {Henderson}, B.~L., \& {Swain}, M. 2019, \apj,
  871, 158

\bibitem[{{Fortney} {et~al.}(2013){Fortney}, {Mordasini}, {Nettelmann},
  {Kempton}, {Greene}, \& {Zahnle}}]{fortney2013framework}
{Fortney}, J.~J., {Mordasini}, C., {Nettelmann}, N., {et~al.} 2013, \apj, 775,
  80

\bibitem[{{Fulton} \& {Petigura}(2018)}]{Fulton2018}
{Fulton}, B.~J., \& {Petigura}, E.~A. 2018, \aj, 156, 264

\bibitem[{{Fulton} {et~al.}(2017){Fulton}, {Petigura}, {Howard}, {Isaacson},
  {Marcy}, {Cargile}, {Hebb}, {Weiss}, {Johnson}, {Morton}, {Sinukoff},
  {Crossfield}, \& {Hirsch}}]{fulton2017}
{Fulton}, B.~J., {Petigura}, E.~A., {Howard}, A.~W., {et~al.} 2017, \aj, 154,
  109

\bibitem[{Furukawa {et~al.}(2019)Furukawa, Chikaraishi, Ohkouchi, Ogawa,
  Glavin, Dworkin, Abe, \& Nakamura}]{Furukawa2019sugarmurchison}
Furukawa, Y., Chikaraishi, Y., Ohkouchi, N., {et~al.} 2019, Proceedings of the
  National Academy of Sciences, doi:10.1073/pnas.1907169116

\bibitem[{{Gao} {et~al.}(2017){Gao}, {Marley}, {Zahnle}, {Robinson}, \&
  {Lewis}}]{gao2017sulfur}
{Gao}, P., {Marley}, M.~S., {Zahnle}, K., {Robinson}, T.~D., \& {Lewis}, N.~K.
  2017, \aj, 153, 139

\bibitem[{{Gautier} {et~al.}(2014){Gautier}, {Carrasco}, {Schmitz-Afonso},
  {Touboul}, {Szopa}, {Buch}, \& {Pernot}}]{gautier2014orbitrap}
{Gautier}, T., {Carrasco}, N., {Schmitz-Afonso}, I., {et~al.} 2014, Earth and
  Planetary Science Letters, 404, 33

\bibitem[{{Gautier} {et~al.}(2016){Gautier}, {Schmitz-Afonso}, {Touboul},
  {Szopa}, {Buch}, \& {Carrasco}}]{gautier2016hplcorbitrap}
{Gautier}, T., {Schmitz-Afonso}, I., {Touboul}, D., {et~al.} 2016, \icarus,
  275, 259

\bibitem[{{Gavilan} {et~al.}(2017){Gavilan}, {Broch}, {Carrasco}, {Fleury}, \&
  {Vettier}}]{gavilan2017}
{Gavilan}, L., {Broch}, L., {Carrasco}, N., {Fleury}, B., \& {Vettier}, L.
  2017, \apjl, 848, L5

\bibitem[{{Gavilan} {et~al.}(2018){Gavilan}, {Carrasco}, {Vr{\o}nning
  Hoffmann}, {Jones}, \& {Mason}}]{gavilan2018organicaerosols}
{Gavilan}, L., {Carrasco}, N., {Vr{\o}nning Hoffmann}, S., {Jones}, N.~C., \&
  {Mason}, N.~J. 2018, \apj, 861, 110

\bibitem[{{Gordon} \& {McBride}(1996)}]{gordonandmcbride}
{Gordon}, S., \& {McBride}, B. 1996, NASA Reference Publication, 1311

\bibitem[{{Grundy} {et~al.}(2018){Grundy}, {Bertrand}, {Binzel}, {Buie},
  {Buratti}, {Cheng}, {Cook}, {Cruikshank}, {Devins}, {Dalle Ore}, {Earle},
  {Ennico}, {Forget}, {Gao}, {Gladstone}, {Howett}, {Jennings}, {Kammer},
  {Lauer}, {Linscott}, {Lisse}, {Lunsford}, {McKinnon}, {Olkin}, {Parker},
  {Protopapa}, {Quirico}, {Reuter}, {Schmitt}, {Singer}, {Spencer}, {Stern},
  {Strobel}, {Summers}, {Weaver}, {Weigle}, {Wong}, {Young}, {Young}, \&
  {Zhang}}]{grundyhazeassurfacematerial}
{Grundy}, W.~M., {Bertrand}, T., {Binzel}, R.~P., {et~al.} 2018, \icarus, 314,
  232

\bibitem[{{Gupta} \& {Schlichting}(2019)}]{GuptaSchlichting2019}
{Gupta}, A., \& {Schlichting}, H.~E. 2019, \mnras, 487, 24

\bibitem[{{Hardegree-Ullman} {et~al.}(2020){Hardegree-Ullman}, {Zink},
  {Christiansen}, {Dressing}, {Ciardi}, \& {Schlieder}}]{hardegreeullman2020}
{Hardegree-Ullman}, K.~K., {Zink}, J.~K., {Christiansen}, J.~L., {et~al.} 2020,
  arXiv e-prints, arXiv:2001.11511

\bibitem[{{Harman} {et~al.}(2013){Harman}, {Kasting}, \&
  {Wolf}}]{harman2013glycoinearlyearth}
{Harman}, C.~E., {Kasting}, J.~F., \& {Wolf}, E.~T. 2013, Origins of Life and
  Evolution of the Biosphere, 43, 77

\bibitem[{{Hartwick} {et~al.}(2019){Hartwick}, {Toon}, \&
  {Heavens}}]{hartwickmarsccn2019}
{Hartwick}, V.~L., {Toon}, O.~B., \& {Heavens}, N.~G. 2019, Nature Geoscience,
  12, 516

\bibitem[{{He} {et~al.}(2017){He}, {H{\"o}rst}, {Riemer}, {Sebree}, {Pauley},
  \& {Vuitton}}]{he2017co}
{He}, C., {H{\"o}rst}, S.~M., {Riemer}, S., {et~al.} 2017, \apj, 841, L31

\bibitem[{{He} \& {Smith}(2014)}]{he2014solubilitytholin}
{He}, C., \& {Smith}, M.~A. 2014, Icarus, 232, 54

\bibitem[{{He} {et~al.}(2018{\natexlab{a}}){He}, {H{\"o}rst}, {Lewis}, {Yu},
  {Moses}, {Kempton}, {McGuiggan}, {Morley}, {Valenti}, \&
  {Vuitton}}]{he2018particle}
{He}, C., {H{\"o}rst}, S.~M., {Lewis}, N.~K., {et~al.} 2018{\natexlab{a}},
  \apj, 856, L3

\bibitem[{{He} {et~al.}(2018{\natexlab{b}}){He}, {H{\"o}rst}, {Lewis}, {Yu},
  {Moses}, {Kempton}, {Marley}, {McGuiggan}, {Morley}, {Valenti}, \&
  {Vuitton}}]{he2018hazeuvplasma}
---. 2018{\natexlab{b}}, \aj, 156, 38

\bibitem[{{He} {et~al.}(2019){He}, {H{\"o}rst}, {Lewis}, {Moses}, {Kempton},
  {Marley}, {Morley}, {Valenti}, \& {Vuitton}}]{he2019gasphase}
---. 2019, ACS Earth Space Chem, 3, 39

\bibitem[{{Helling}(2019)}]{hellingreview2019}
{Helling}, C. 2019, Annual Review of Earth and Planetary Sciences, 47, 583

\bibitem[{{Hollis} {et~al.}(2000){Hollis}, {Lovas}, \&
  {Jewell}}]{hollis2000sugarism}
{Hollis}, J.~M., {Lovas}, F.~J., \& {Jewell}, P.~R. 2000, \apjl, 540, L107

\bibitem[{{H{\"o}rst}(2011)}]{horstthesis}
{H{\"o}rst}, S.~M. 2011, PhD thesis, PhD Thesis, University of
  Arizona.~2011.~Advisor: Yelle, Roger V.

\bibitem[{{H{\"o}rst}(2017)}]{horsttitanreview}
---. 2017, Journal of Geophysical Research (Planets), 122, 432

\bibitem[{{H{\"o}rst} {et~al.}(2018{\natexlab{a}}){H{\"o}rst}, {He}, {Ugelow},
  {Jellinek}, {Pierrehumbert}, \& {Tolbert}}]{horst2018earlyearth}
{H{\"o}rst}, S.~M., {He}, C., {Ugelow}, M.~S., {et~al.} 2018{\natexlab{a}},
  \apj, 858, 119

\bibitem[{{H{\"o}rst} \& {Tolbert}(2014)}]{horsttolbert2014}
{H{\"o}rst}, S.~M., \& {Tolbert}, M.~A. 2014, The Astrophysical Journal, 781,
  53

\bibitem[{{H{\"o}rst} {et~al.}(2018{\natexlab{b}}){H{\"o}rst}, {Yoon},
  {Ugelow}, {Parker}, {Li}, {de Gouw}, \&
  {Tolbert}}]{horst2018titanlabparticlegas}
{H{\"o}rst}, S.~M., {Yoon}, Y.~H., {Ugelow}, M.~S., {et~al.}
  2018{\natexlab{b}}, Icarus, 301, 136

\bibitem[{{H{\"o}rst} {et~al.}(2012){H{\"o}rst}, {Yelle}, {Buch}, {Carrasco},
  {Cernogora}, {Dutuit}, {Quirico}, {Sciamma-O'Brien}, {Smith}, {Somogyi}, \&
  {Szopa}}]{horst2012formation}
{H{\"o}rst}, S.~M., {Yelle}, R.~V., {Buch}, A., {et~al.} 2012, Astrobiology,
  12, 809

\bibitem[{{H{\"o}rst} {et~al.}(2018{\natexlab{c}}){H{\"o}rst}, {He}, {Lewis},
  {Kempton}, {Marley}, {Morley}, {Moses}, {Valenti}, \&
  {Vuitton}}]{horst2018production}
{H{\"o}rst}, S.~M., {He}, C., {Lewis}, N.~K., {et~al.} 2018{\natexlab{c}},
  Nature Astronomy, 2

\bibitem[{{Howe} \& {Burrows}(2012)}]{howe2012theoretical}
{Howe}, A.~R., \& {Burrows}, A.~S. 2012, \apj, 756, 176

\bibitem[{{Hu} {et~al.}(2005){Hu}, {Noll}, {Li}, {Makarov}, {Hardman}, \&
  {Graham Cooks}}]{huorbitrap2005}
{Hu}, Q., {Noll}, R.~J., {Li}, H., {et~al.} 2005, Journal of Mass Spectrometry,
  40, 430

\bibitem[{{Hu} \& {Seager}(2014)}]{hu2014photochemistry}
{Hu}, R., \& {Seager}, S. 2014, \apj, 784, 63

\bibitem[{{Imanaka} \& {Smith}(2007)}]{Imanaka2007}
{Imanaka}, H., \& {Smith}, M.~A. 2007, \grl, 34, L02204

\bibitem[{{Jalbout} {et~al.}(2007){Jalbout}, {Abrell}, {Adamowicz}, {Polt},
  {Apponi}, \& {Ziurys}}]{jalbout2007formosegasphase}
{Jalbout}, A.~F., {Abrell}, L., {Adamowicz}, L., {et~al.} 2007, Astrobiology,
  7, 433

\bibitem[{{Kawashima} \& {Ikoma}(2019)}]{kawashima2019hazemodel}
{Kawashima}, Y., \& {Ikoma}, M. 2019, \apj, 877, 109

\bibitem[{Kim {et~al.}(2003)Kim, Kramer, \& Hatcher}]{Kim2003VKoverview}
Kim, S., Kramer, R.~W., \& Hatcher, P.~G. 2003, Analytical Chemistry, 75, 5336,
  pMID: 14710810

\bibitem[{{Knutson} {et~al.}(2014){Knutson}, {Benneke}, {Deming}, \&
  {Homeier}}]{knutson2014featureless}
{Knutson}, H.~A., {Benneke}, B., {Deming}, D., \& {Homeier}, D. 2014, Nature,
  505, 66

\bibitem[{{Kopetzki} \& {Antonietti}(2011)}]{kopetzki2011hydrothermalformose}
{Kopetzki}, D., \& {Antonietti}, M. 2011, New Journal of Chemistry, 35, 1787

\bibitem[{{Kreidberg} {et~al.}(2014){Kreidberg}, {Bean}, {D{\'e}sert},
  {Benneke}, {Deming}, {Stevenson}, {Seager}, {Berta-Thompson}, {Seifahrt}, \&
  {Homeier}}]{kreidberg2014clouds}
{Kreidberg}, L., {Bean}, J.~L., {D{\'e}sert}, J.-M., {et~al.} 2014, Nature,
  505, 69

\bibitem[{{Kreidberg} {et~al.}(2019){Kreidberg}, {Koll}, {Morley}, {Hu},
  {Schaefer}, {Deming}, {Stevenson}, {Dittmann}, {Vanderburg}, {Berardo},
  {Guo}, {Stassun}, {Crossfield}, {Charbonneau}, {Latham}, {Loeb}, {Ricker},
  {Seager}, \& {Vand erspek}}]{Kreidberg2019}
{Kreidberg}, L., {Koll}, D. D.~B., {Morley}, C., {et~al.} 2019, \nat, 573, 87

\bibitem[{{Kwok}(2016)}]{sun2016organicsreview}
{Kwok}, S. 2016, \aapr, 24, 8

\bibitem[{{Leconte} {et~al.}(2015){Leconte}, {Forget}, \&
  {Lammer}}]{Leconte2015}
{Leconte}, J., {Forget}, F., \& {Lammer}, H. 2015, Experimental Astronomy, 40,
  449

\bibitem[{{Lehmer} \& {Catling}(2017)}]{LehmerCatling2017}
{Lehmer}, O.~R., \& {Catling}, D.~C. 2017, \apj, 845, 130

\bibitem[{{Lopez} \& {Fortney}(2014)}]{lopezfortney2014}
{Lopez}, E.~D., \& {Fortney}, J.~J. 2014, \apj, 792, 1

\bibitem[{{Loyd} {et~al.}(2020){Loyd}, {Shkolnik}, {Schneider},
  {Richey-Yowell}, {Barman}, {Peacock}, \& {Pagano}}]{Loyd2020}
{Loyd}, R.~O.~P., {Shkolnik}, E.~L., {Schneider}, A.~C., {et~al.} 2020, \apj,
  890, 23

\bibitem[{{Lu} \& {Freeland}(2006)}]{lufreelandaminoacidlists}
{Lu}, Y., \& {Freeland}, S. 2006, Astrobiology, 6, 606

\bibitem[{{Maillard} {et~al.}(2018){Maillard}, {Carrasco}, {Schmitz-Afonso},
  {Gautier}, \& {Afonso}}]{maillard2018soluble}
{Maillard}, J., {Carrasco}, N., {Schmitz-Afonso}, I., {Gautier}, T., \&
  {Afonso}, C. 2018, Earth and Planetary Science Letters, 495, 185

\bibitem[{{Marley} {et~al.}(2013){Marley}, {Ackerman}, {Cuzzi}, \&
  {Kitzmann}}]{marley2013review}
{Marley}, M.~S., {Ackerman}, A.~S., {Cuzzi}, J.~N., \& {Kitzmann}, D. 2013,
  ``Clouds and Hazes in Exoplanet Atmospheres'' (Tucson: University of Arizona
  Press), 367--391

\bibitem[{{Meinert} {et~al.}(2016){Meinert}, {Myrgorodska}, {de Marcellus},
  {Buhse}, {Nahon}, {Hoffmann}, {d'Hendecourt}, \&
  {Meierhenrich}}]{meinert2016riboseice}
{Meinert}, C., {Myrgorodska}, I., {de Marcellus}, P., {et~al.} 2016, Science,
  352, 208

\bibitem[{{Miller}(1953)}]{miller1953}
{Miller}, S.~L. 1953, Science, 117, 528

\bibitem[{{Miller} \& {Urey}(1959)}]{millerurey1959}
{Miller}, S.~L., \& {Urey}, H.~C. 1959, Science, 130, 245

\bibitem[{{Miller-Ricci Kempton} {et~al.}(2012){Miller-Ricci Kempton},
  {Zahnle}, \& {Fortney}}]{millerriccikempton2012chemistry}
{Miller-Ricci Kempton}, E., {Zahnle}, K., \& {Fortney}, J.~J. 2012, \apj, 745,
  3

\bibitem[{{Morley} {et~al.}(2013){Morley}, {Fortney}, {Kempton}, {Marley},
  {Visscher}, \& {Zahnle}}]{morley2013quantatively}
{Morley}, C.~V., {Fortney}, J.~J., {Kempton}, E. M.~R., {et~al.} 2013, \apj,
  775, 33

\bibitem[{{Morley} {et~al.}(2015){Morley}, {Fortney}, {Marley}, {Zahnle},
  {Line}, {Kempton}, {Lewis}, \& {Cahoy}}]{morley2015thermal}
{Morley}, C.~V., {Fortney}, J.~J., {Marley}, M.~S., {et~al.} 2015, \apj, 815,
  110

\bibitem[{{Morley} {et~al.}(2017){Morley}, {Knutson}, {Line}, {Fortney},
  {Thorngren}, {Marley}, {Teal}, \& {Lupu}}]{Morley2017gj436b}
{Morley}, C.~V., {Knutson}, H., {Line}, M., {et~al.} 2017, \aj, 153, 86

\bibitem[{{Moses} {et~al.}(2013){Moses}, {Line}, {Visscher}, {Richardson},
  {Nettelmann}, {Fortney}, {Barman}, {Stevenson}, \&
  {Madhusudhan}}]{moses2013equilibriumcomposition}
{Moses}, J.~I., {Line}, M.~R., {Visscher}, C., {et~al.} 2013, \apj, 777, 34

\bibitem[{{Neish} {et~al.}(2010){Neish}, {Somogyi}, \&
  {Smith}}]{Neish2010massspectitanhydrolysis}
{Neish}, C.~D., {Somogyi}, {\'A}., \& {Smith}, M.~A. 2010, Astrobiology, 10,
  337

\bibitem[{{Nuevo} {et~al.}(2018){Nuevo}, {Cooper}, \&
  {Sandford}}]{nuevo2018sugarice}
{Nuevo}, M., {Cooper}, G., \& {Sandford}, S.~A. 2018, Nature Communications, 9,
  5276

\bibitem[{{Owen} \& {Wu}(2016)}]{OwenWu2016}
{Owen}, J.~E., \& {Wu}, Y. 2016, \apj, 817, 107

\bibitem[{{Peacock} {et~al.}(2019){Peacock}, {Barman}, {Shkolnik},
  {Hauschildt}, \& {Baron}}]{peacock2019uvradiation}
{Peacock}, S., {Barman}, T., {Shkolnik}, E.~L., {Hauschildt}, P.~H., \&
  {Baron}, E. 2019, \apj, 871, 235

\bibitem[{{Perry} {et~al.}(2008){Perry}, {Cooks}, \&
  {Noll}}]{perryorbitrap2008}
{Perry}, R.~H., {Cooks}, R.~G., \& {Noll}, R.~J. 2008, Mass Spectrometry
  Reviews, 27, 661

\bibitem[{{Pestunova} {et~al.}(2005){Pestunova}, {Simonov}, {Snytnikov},
  {Stoyanovsky}, \& {Parmon}}]{Pestunova2005formoseUVrad}
{Pestunova}, O., {Simonov}, A., {Snytnikov}, V., {Stoyanovsky}, V., \&
  {Parmon}, V. 2005, Advances in Space Research, 36, 214

\bibitem[{{Ranjan} {et~al.}(2017){Ranjan}, {Wordsworth}, \&
  {Sasselov}}]{ranjan2017uvmdwarf}
{Ranjan}, S., {Wordsworth}, R., \& {Sasselov}, D.~D. 2017, \apj, 843, 110

\bibitem[{{Ricker} {et~al.}(2014){Ricker}, {Winn}, {Vanderspek}, {Latham},
  {Bakos}, {Bean}, {Berta-Thompson}, {Brown}, {Buchhave}, {Butler}, {Butler},
  {Chaplin}, {Charbonneau}, {Christensen-Dalsgaard}, {Clampin}, {Deming},
  {Doty}, {De Lee}, {Dressing}, {Dunham}, {Endl}, {Fressin}, {Ge}, {Henning},
  {Holman}, {Howard}, {Ida}, {Jenkins}, {Jernigan}, {Johnson}, {Kaltenegger},
  {Kawai}, {Kjeldsen}, {Laughlin}, {Levine}, {Lin}, {Lissauer}, {MacQueen},
  {Marcy}, {McCullough}, {Morton}, {Narita}, {Paegert}, {Palle}, {Pepe},
  {Pepper}, {Quirrenbach}, {Rinehart}, {Sasselov}, {Sato}, {Seager},
  {Sozzetti}, {Stassun}, {Sullivan}, {Szentgyorgyi}, {Torres}, {Udry}, \&
  {Villasenor}}]{rickerTESS}
{Ricker}, G.~R., {Winn}, J.~N., {Vanderspek}, R., {et~al.} 2014, Proceedings of
  the SPIE, 9143, 914320

\bibitem[{{Rimmer} {et~al.}(2018){Rimmer}, {Xu}, {Thompson}, {Gillen},
  {Sutherland}, \& {Queloz}}]{rimmer2018rna}
{Rimmer}, P.~B., {Xu}, J., {Thompson}, S.~J., {et~al.} 2018, Science Advances,
  4, eaar3302

\bibitem[{Ruf {et~al.}(2018)Ruf, Le~Sergeant~d'Hendecourt, \&
  Schmitt-Kopplin}]{Ruf2018VKastro}
Ruf, A., Le~Sergeant~d'Hendecourt, L., \& Schmitt-Kopplin, P. 2018, Life, 8, 18

\bibitem[{{Sarker} {et~al.}(2003){Sarker}, {Somogyi}, {Lunine}, \&
  {Smith}}]{Sarkertitanmassspec}
{Sarker}, N., {Somogyi}, A., {Lunine}, J.~I., \& {Smith}, M.~A. 2003,
  Astrobiology, 3, 719

\bibitem[{{Schaefer} {et~al.}(2012){Schaefer}, {Lodders}, \&
  {Fegley}}]{schaefer2012}
{Schaefer}, L., {Lodders}, K., \& {Fegley}, B. 2012, \apj, 755, 41

\bibitem[{Schwartz(2007)}]{schwartz2007yields}
Schwartz, A. 2007, Chemistry \& Biodiversity, 4, 656

\bibitem[{{Schwartz} {et~al.}(1984){Schwartz}, {Voet}, \& {van der
  Veen}}]{schwartz1984hcnprebiotic}
{Schwartz}, A.~W., {Voet}, A.~B., \& {van der Veen}, M. 1984, Origins of Life,
  14, 91

\bibitem[{{Sebree} {et~al.}(2018){Sebree}, {Roach}, {Shipley}, {He}, \&
  {H{\"o}rst}}]{sebree2018}
{Sebree}, J.~A., {Roach}, M.~C., {Shipley}, E.~R., {He}, C., \& {H{\"o}rst},
  S.~M. 2018, \apj, 865, 133

\bibitem[{{Sing} {et~al.}(2016){Sing}, {Fortney}, {Nikolov}, {Wakeford},
  {Kataria}, {Evans}, {Aigrain}, {Ballester}, {Burrows}, {Deming},
  {D{\'e}sert}, {Gibson}, {Henry}, {Huitson}, {Knutson}, {Lecavelier Des
  Etangs}, {Pont}, {Showman}, {Vidal-Madjar}, {Williamson}, \&
  {Wilson}}]{sing2016continuum}
{Sing}, D.~K., {Fortney}, J.~J., {Nikolov}, N., {et~al.} 2016, Nature, 529, 59

\bibitem[{Somogyi {et~al.}(2016)Somogyi, Thissen, Orthous-Daunay, \&
  Vuitton}]{somogyi2016}
Somogyi, A., Thissen, R., Orthous-Daunay, F.-R., \& Vuitton, V. 2016,
  International Journal of Molecular Sciences, 17, 439

\bibitem[{{Trainer} {et~al.}(2012){Trainer}, {Jimenez}, {Yung}, {Toon}, \&
  {Tolbert}}]{trainer2012nitrogen}
{Trainer}, M.~G., {Jimenez}, J.~L., {Yung}, Y.~L., {Toon}, O.~B., \& {Tolbert},
  M.~A. 2012, Astrobiology, 12, 315

\bibitem[{{Trainer} {et~al.}(2006){Trainer}, {Pavlov}, {Dewitt}, {Jimenez},
  {McKay}, {Toon}, \& {Tolbert}}]{TrainerPNAS2006}
{Trainer}, M.~G., {Pavlov}, A.~A., {Dewitt}, H.~L., {et~al.} 2006, Proceedings
  of the National Academy of Science, 103, 18035

\bibitem[{{Tsiaras} {et~al.}(2019){Tsiaras}, {Waldmann}, {Tinetti}, {Tennyson},
  \& {Yurchenko}}]{tsiaras2019k2}
{Tsiaras}, A., {Waldmann}, I.~P., {Tinetti}, G., {Tennyson}, J., \&
  {Yurchenko}, S.~N. 2019, Nature Astronomy, 3, 1086

\bibitem[{{Vuitton} {et~al.}(2014){Vuitton}, {Briois}, \&
  {Makarov}}]{vuitton2014_massspec_inspace}
{Vuitton}, V., {Briois}, C., \& {Makarov}, A.~e. 2014, in 40th COSPAR
  Scientific Assembly, Vol.~40, B0.6--2--14

\bibitem[{{Vuitton} {et~al.}(2019){Vuitton}, {Yelle}, {Klippenstein},
  {H{\"o}rst}, \& {Lavvas}}]{vuitton2019}
{Vuitton}, V., {Yelle}, R.~V., {Klippenstein}, S.~J., {H{\"o}rst}, S.~M., \&
  {Lavvas}, P. 2019, \icarus, 324, 120

\bibitem[{{Vuitton} {et~al.}(2010){Vuitton}, {Bonnet}, {Frisari}, {Thissen},
  {Quirico}, {Dutuit}, {Schmitt}, {Le Roy}, {Fray}, {Cottin},
  {Sciamma-O'Brien}, {Carrasco}, \& {Szopa}}]{vuitton2010HCN}
{Vuitton}, V., {Bonnet}, J.-Y., {Frisari}, M., {et~al.} 2010, Faraday
  Discussions, 147, 495

\bibitem[{{Wakeford} {et~al.}(2019){Wakeford}, {Wilson}, {Stevenson}, \&
  {Lewis}}]{wakeford2019rnaas}
{Wakeford}, H.~R., {Wilson}, T.~J., {Stevenson}, K.~B., \& {Lewis}, N.~K. 2019,
  Research Notes of the American Astronomical Society, 3, 7

\bibitem[{{Youngblood} {et~al.}(2017){Youngblood}, {France}, {Loyd}, {Brown},
  {Mason}, {Schneider}, {Tilley}, {Berta-Thompson}, {Buccino}, {Froning},
  {Hawley}, {Linsky}, {Mauas}, {Redfield}, {Kowalski}, {Miguel}, {Newton},
  {Rugheimer}, {Segura}, {Roberge}, \&
  {Vieytes}}]{youngblood2017uvflaresmdwarfs}
{Youngblood}, A., {France}, K., {Loyd}, R.~O.~P., {et~al.} 2017, \apj, 843, 31

\bibitem[{{Zhang} {et~al.}(2017){Zhang}, {Strobel}, \&
  {Imanaka}}]{zhangpluto2017}
{Zhang}, X., {Strobel}, D.~F., \& {Imanaka}, H. 2017, \nat, 551, 352

\end{thebibliography}
\bibliographystyle{aasjournal}

\end{document}